\documentclass{aa}
\usepackage{graphicx}
\usepackage[varg]{txfonts}
\usepackage{natbib}
\usepackage{gensymb}
\usepackage{multirow}
\usepackage{color}
\usepackage[normalem]{ulem}
\usepackage[]{siunitx}
\usepackage{rotating}

\bibpunct{(}{)}{;}{a}{}{,} 
\usepackage{amstext}

\begin{document}

\title{
High-resolution near-infrared spectroscopy of globular cluster and field stars toward the Galactic bulge 
}

\author{Dongwook Lim\inst{1}
  \and Andreas J. Koch-Hansen\inst{1}
  \and Sang-Hyun Chun\inst{2}
  \and Seungsoo Hong\inst{3} 
  \and Young-Wook Lee\inst{3}
  }

\authorrunning{D. Lim et al.}
\titlerunning{High-resolution NIR spectroscopy of GC and field stars toward the Galactic bulge }

\institute{Zentrum f\"ur Astronomie der Universit\"at Heidelberg, Astronomisches Rechen-Institut, M\"onchhofstr. 12-14, 69120 Heidelberg, Germany, 
  \email{dongwook.lim@uni-heidelberg.de}
  \and Korea Astronomy and Space Science Institute, 776 Daedeokdae-ro, Yuseong-gu, Daejeon 34055, Republic of Korea
  \and Center for Galaxy Evolution Research \& Department of Astronomy, Yonsei University, Seoul 03722, Republic of Korea
  }

\date{Received / Accepted }

\abstract {
Globular clusters (GCs) play an important role in the formation and evolution of the Milky Way.
New candidates are continuously found, particularly in the high-extinction low-latitude regions of the bulge,
although their existence and properties have yet to be verified. 
In order to investigate the new GC candidates, we performed high-resolution near-infrared spectroscopy of stars toward the Galactic bulge using the Immersion Grating Infrared Spectrometer (IGRINS) instrument at the Gemini-South telescope. 
We selected 15 and 10 target stars near Camargo~1103 and Camargo~1106, respectively, which have recently been reported as metal-poor GC candidates in the bulge. 
In contrast to the classical approaches used in optical spectroscopy, we determined stellar parameters from a combination of line-depth ratios and the equivalent width of a CO line. 
The stellar parameters of the stars follow the common trends of nearby APOGEE sample stars in a similar magnitude range.
We also determined the abundances of Fe, Na, Mg, Al, Si, S, K, Ca, Ti, Cr, Ni, and Ce through spectrum synthesis. 
There is no clear evidence of a grouping in radial velocity - metallicity space that would indicate the characterization of either object as metal-poor GCs.
This result emphasizes the necessity of follow-up spectroscopy for new GC candidates toward the bulge, although we cannot completely rule out a low probability that we only observed nonmember stars.
We also note discrepancies between the abundances of Al, Ca, and Ti when derived from the H- versus the K-band spectra. 
Although the cause of this discrepancy is not clear, the effects of atmosphere parameters or nonlocal thermodynamic equilibrium are discussed.
Our approach and results demonstrate that IGRINS spectroscopy is a useful tool for studying the chemical properties of stars toward the Galactic bulge with a statistical uncertainty in [Fe/H] of $\sim$0.03~dex, while the systematic error through uncertainties of atmospheric parameter determination, at $\sim$0.14~dex, is slightly larger than in measurements from optical spectroscopy.
} 

\maketitle


\section{Introduction} \label{sec:intro}
Detailed chemical abundance patterns of stars are key to understanding the formation and evolution of the Milky Way (MW) because they contain significant information about their birthplace and evolution processes \citep{Freeman2002}. 
The chemical tagging approach is now widely employed in Galactic archaeology based on extensive spectroscopic surveys, such as the Apache Point Observatory Galactic Evolution Experiment \citep[APOGEE;][]{Majewski2017} and the Galactic Archaeology with HERMES \citep[GALAH;][]{DeSilva2015} surveys. 
The recent data release of APOGEE (DR16; \citealt{Ahumada2020}), for example, provides chemical abundances {for 26 species}, and radial velocities and atmosphere parameters for more than 430,000 stars obtained from the near-infrared (NIR) H-band spectra. 

One of the most active research areas for the MW is the search for new globular clusters (GCs) in the Galactic bulge and the examination of their chemical and kinematic properties. Galactic GCs play an important role in the formation and evolution history of the MW because they have a fossil record of their birthplace.
Compared to the halo, however, only a small number of GCs were found at the low latitudes of the MW bulge, and their nature is little studied because of the high extinction and confusion with the disks and overlapping inner halo.
Recent large NIR photometric surveys, such as the Vista Variables in the Via Lactea \citep[VVV;][]{Minniti2010}, reported an increasing number of new stellar cluster candidates in the bulge \citep[e.g.,][]{Gran2019, Minniti2019, Minniti2021}, although further confirmation by follow-up spectroscopy is required.
\citet{Camargo2018} and \citet{Camargo2019} also discovered eight new GC candidates in the low-latitude field of the bulge ($|b|$ $\sim$ 2$\degree$) using photometry from the Two Micron All Sky Survey (2MASS), Wide-field Infrared Survey Explorer (WISE), VVV, and Gaia-DR2.
In particular, they suggested that these candidates are rare GCs in the bulge showing old ages ($>$ 12 Gyr) and metal-poor properties ([Fe/H] $<$ $-$1.5 dex) according to the isochrone fitting \citep[see][]{Bica2016}.  
If this is the case, these sources are crucial for studying the bulge because old and metal-poor GCs could be evidence for a classical bulge component that formed in the early stage of the MW formation \citep[see, e.g.,][]{Nataf2017, Lee2019}.
In addition, their chemical properties can provide important information for the connection to the bulge field stars that are thought to be GC progeny \citep{Schiavon2017, Fernandez-Trincado2022}.
Even if these GCs belong to the inner halo, they are valuable for the study of the kinematics and accretion history of the MW as elusive GCs in the low-latitude field.
However, the above approaches require follow-up spectroscopy for the chemical abundances and line-of-sight velocity. 

Although large spectroscopic survey data are widely employed in various fields, individual high-resolution spectroscopic observations are still required to determine the accurate abundances of various elements for specific systems or individual stars. 
In particular, stars or stellar clusters toward the Galactic bulge are hard to study in surveys due to crowding and/or severe interstellar extinction \citep[e.g.,][]{Koch2017ESO,Gonzalez2020}. 
The Immersion Grating Infrared Spectrometer \citep[IGRINS;][]{Mace2018}, which is currently mounted on the Gemini-South telescope, offers a strong advantage for investigating the detailed chemical properties of stars toward the Galactic bulge. 
Its high spectral resolution (R $\sim$ 45,000) in NIR region (here, the H- and K-bands, which are conveniently insensitive to extinction), combined with the high light-gathering power of an 8.1m telescope, enable obtaining high-quality spectra within reasonable exposure times. 
Thus, this instrument is efficient in observing faint stars in the high-extinction region, where it is more difficult to obtain high-quality spectra from optical spectroscopy or surveys \citep[cf.][]{Bensby2019}. 
 
High-resolution NIR stellar spectroscopy often needs to rely on approaches that differ from standard techniques in the optical.
For instance, it is of limited use for determining atmosphere parameters through standard methods such as an equivalent width (EW) analysis due to the contamination with nearby molecular bands and lines that hamper precise EW measurements. 
Furthermore, the Fe\,{\sc ii} lines that needed to determine spectroscopic surface gravities are rare in the NIR region.
Photometric and astrometric information for stars toward the bulge is also notorious for stellar parameter determination because of the high extinction, partly differential reddening, and crowding. 
Despite these obstacles, an increasing number of chemical abundance studies has been performed in the NIR region \citep[e.g.,][]{D'Orazi2018, Sameshima2018, Ishikawa2022}.
In the case of APOGEE, atmosphere parameters and chemical abundances are determined by spectrum fitting \citep{Jonsson2020}. 
\citet{Fukue2015} applied the line-depth ratios to derive effective temperatures from NIR spectra, and \citet{Park2018} suggested empirical relations between the temperature and EW of selected absorption lines. 
In addition, detailed chemical abundances of red giant and horizontal branch stars were measured through spectral synthesis from IGRINS data before \citep[e.g.,][]{Afsar2018, BocekTopcu2020}, although these studies adopted atmosphere parameters from already existing high-resolution optical spectroscopy. 

In this regard, when the method for determining the atmosphere parameter and abundance measurement from the NIR spectrum is established,  IGRINS is a powerful instrument for studying the formation and evolution of stars in the bulge. 
For example, the accurate metallicity and radial velocity of stars will clarify whether the recently reported candidates from NIR photometric surveys are actual GCs.
Then, multiple stellar populations in the GCs can be examined in terms of Na-O and Mg-Al anticorrelations \citep{Carretta2009, Bastian2018}, which are available to be measured from the IGRINS spectra. 
In addition, the accessibility to the $\alpha$-elements (Mg, Si, S, and Ca) allows us to trace  stars with potential accretion origins because stars formed in a low-mass environment with a low star-forming efficiency have a low [$\alpha$/Fe] abundance ratio \citep{Nissen2010}. 

In order to use the IGRINS data for stars toward the bulge and examine their validity, we observed 25 stars located in the low-latitude field of the MW ($|b|$ $\sim$ 2$\degree$).
This paper is organized as follows. 
We describe the target selection, observation, and data reduction process in Section~\ref{sec:obs}. 
The atmosphere parameters are determined and the spectra are analyzed in Section~\ref{sec:analysis}.
Finally, in Sections~\ref{sec:result} and \ref{sec:discussion}, we present our results and discuss the validity and limitations of spectroscopy using IGRINS. 


\section{Observations and data reduction} \label{sec:obs}
We have selected target stars in the close vicinity
of the bulge GC candidates Camargo~1103 and Camargo~1106, 
which are the brightest objects in the list of eight such newly discovered systems by \citet{Camargo2018} and \citet{Camargo2019}. 
Our aim was to spectroscopically confirm these GC candidates, and to perform a validity check of  IGRINS observations for stars toward the bulge. 

\begin{figure}
\centering
   \includegraphics[width=0.48\textwidth]{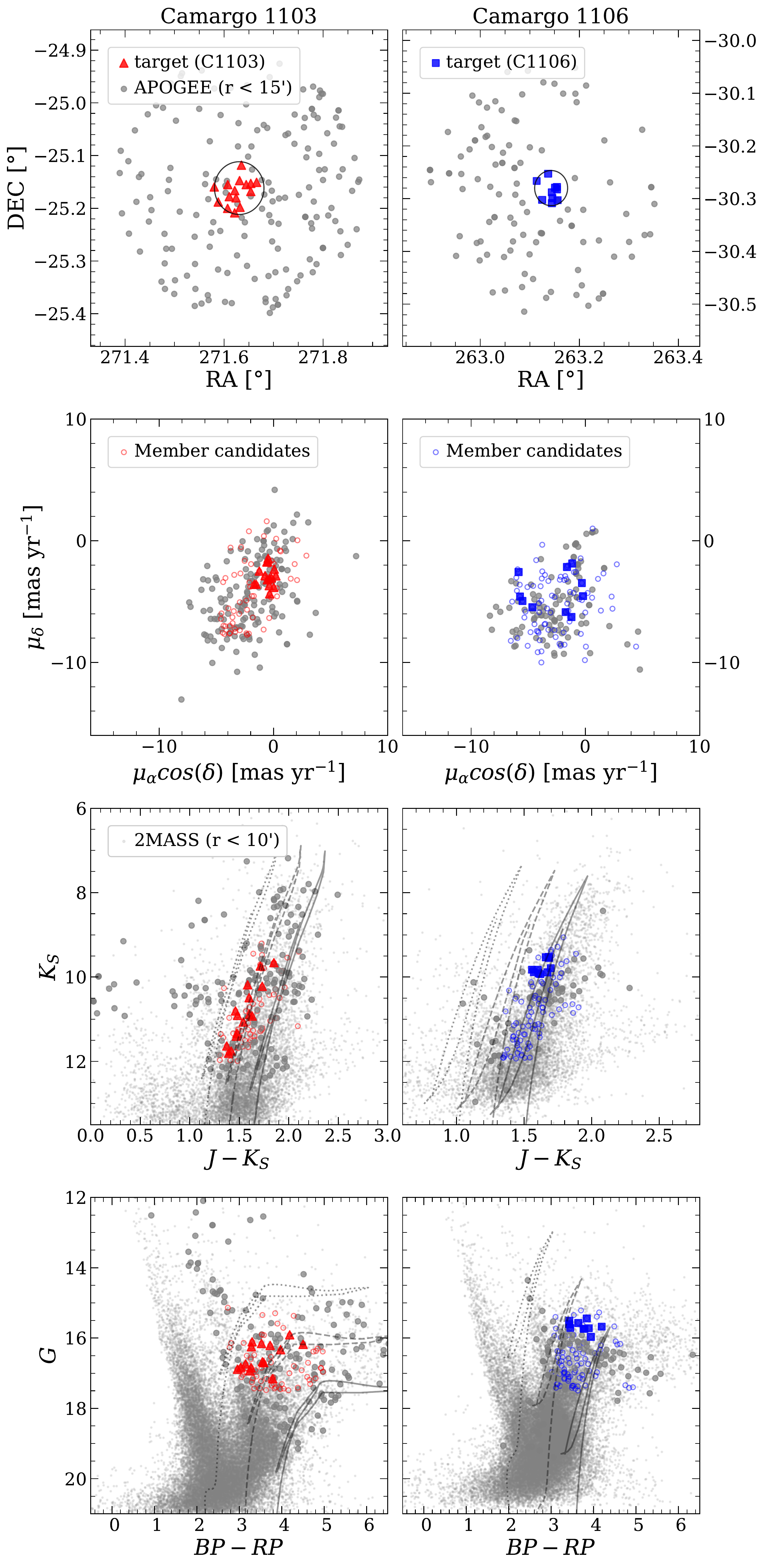}
     \caption{
     Target stars (filled red triangles and filled blue squares) in equatorial coordinates and proper motion, 
     taken from Gaia EDR3 (top half of the panels), and in the 2MASS and Gaia CMDs (bottom two panels). 
     The open red and blue circles indicate member candidates selected based on color and magnitude criteria (see text).
     Our data are shown together with nearby APOGEE stars within a circle of 15$\arcmin$ radius around Camargo~1103 and 1106 (gray circles).
     The black circles in the top panels are circles with radii of 3$\arcmin$ (Camargo~1103) and 2$\arcmin$ (Camargo~1106), in which \citet{Camargo2018} selected member stars.  
     The gray dots in the CMDs are full stars of 2MASS catalog ($<$ 10$\arcmin$ radius from the center).
     The dotted, dashed, and solid lines in the CMDs are BaSTI isochrones of 13.5~Gyr and [Fe/H] = $-$1.80~dex for Camargo~1103 and 12.5~Gyr and [Fe/H]  = $-$1.5~dex for Camargo~1106 \citep{Pietrinferni2021} with E(B-V) = 1.0, 1.5, and 2.0 mag, respectively, to account for the uncertain reddening.
     }
     \label{fig:cmd}
\end{figure}

\begin{table*}
\caption{Target information}
\label{tab:info} 
\centering                                    
\begin{tabular}{l c c c c r r r r}  
\hline\hline 
\multirow{2}{*}{ID}     & $\alpha$ (J2000)      & $\delta$ (J2000)      & $G$             & $K_{S}$               & Exposure $^{a}$       & S/N$_{\rm H-band}$        & S/N$_{\rm K-band}$    & RV$_{helio}$ \\
                                & [hh:mm:ss]            & [dd:mm:ss]            & [mag]           & [mag]         & [second]              & [pixel$^{-1}$]                        & [pixel$^{-1}$]                  & [km~s$^{-1}$] \\
\hline                           
C1103-7052 & 18:06:32.41 & -25:07:05.32 & 16.16 & 9.66 & 320 & 126 & 110 & $-$72.99 $\pm$ 0.29 \\
C1103-8510 & 18:06:31.52 & -25:08:50.97 & 16.18 & 10.49 & 1100 & 118 & 81 & 111.11 $\pm$ 0.33 \\
C1103-9023 & 18:06:39.77 & -25:09:02.38 & 16.13 & 10.80 & 1200 & 110 & 75 & $-$27.19 $\pm$ 0.43 \\
C1103-9100 & 18:06:36.94 & -25:09:10.09 & 16.32 & 10.22 & 800  & 107 & 75 & 174.20 $\pm$ 45 \\
C1103-9180 & 18:06:25.79 & -25:09:18.12 & 16.87 & 11.64 & 1800 & 131 & 74 & 46.42 $\pm$ 0.65 \\
C1103-9181 & 18:06:34.74 & -25:09:18.09 & 15.91 &  9.74 & 520  & 110 & 79 & $-$144.40 $\pm$ 0.23 \\
C1103-9341 & 18:06:19.25 & -25:09:34.12 & 16.67 & 10.89 & 1280 & 128 & 89 & $-$38.27 $\pm$ 0.23 \\
C1103-9590 & 18:06:29.18 & -25:09:59.13 & 16.77 & 11.33 & 1800 & 228 & 136 & $-$40.94 $\pm$ 0.50 \\
C1103-10048 & 18:06:36.95 & -25:10:04.92 & 16.73 & 10.93 & 1600 & 128 & 78 & $-$68.63 $\pm$ 0.30 \\
C1103-10405 & 18:06:26.52 & -25:10:40.41 & 16.92 & 11.75 & 1800 & 118 & 66 & $-$57.99 $\pm$ 0.35 \\
C1103-10484 & 18:06:29.81 & -25:10:48.47 & 16.25 & 10.91 & 1000 & 114 & 76 & 31.62 $\pm$ 0.38 \\
C1103-11170 & 18:06:21.18 & -25:11:17.04 &  16.96 & 11.40 & 1800 & 130 & 78 & 176.30 $\pm$ 0.31 \\
C1103-11530 & 18:06:31.74 & -25:11:53.04 &  16.84 & 11.81 & 1600 & 111 & 73 & $-$36.98 $\pm$ 0.46 \\
C1103-12006 & 18:06:25.72 & -25:12:00.36 & 16.21 & 10.18 & 780  & 100 & 67 & $-$112.95 $\pm$ 0.31 \\
C1103-12308 & 18:06:29.13 & -25:12:30.73 & 17.16 & 11.07 & 1800 & 108 & 65 & $-$94.11 $\pm$ 0.40 \\
\hline
C1106-15096 & 17:32:32.91 & -30:15:09.65 & 15.76 & 9.79 & 540 & 99 & 80 & $-$94.80 $\pm$ 0.79 \\
C1106-15592 & 17:32:27.36 & -30:15:59.44 & 15.77 & 9.89 & 600 & 102 & 75 & 166.89 $\pm$ 0.26 \\
C1106-16402 & 17:32:37.17 & -30:16:39.99 & 15.74 & 9.52 & 320 & 140 & 112 & 138.29 $\pm$ 0.41 \\
C1106-16451 & 17:32:36.27 & -30:16:45.28 & 15.69 & 9.55 & 440 & 154 & 131 & 122.28 $\pm$ 0.39 \\ 
C1106-16566 & 17:32:37.20 & -30:16:56.69 & 15.60 & 9.83 & 560 & 80 & 64 & $-$154.51 $\pm$ 1.08 \\
C1106-17160 & 17:32:34.62 & -30:17:16.18 & 15.63 & 9.93 & 640 & 126 & 101 & $-$90.67 $\pm$ 0.52 \\
C1106-17565 & 17:32:34.98 & -30:17:56.63 & 15.99 & 9.91 & 640 & 114 & 81 & 50.64 $\pm$ 0.31 \\
C1106-18086 & 17:32:30.01 & -30:18:08.65 & 15.73 & 9.88 & 600 & 105 & 74 & 208.56 $\pm$ 0.46 \\
C1106-18101 & 17:32:37.44 & -30:18:10.35 & 15.54 & 9.83 & 560 & 124 & 92 & 6.62 $\pm$ 0.45 \\
C1106-18303 & 17:32:34.75 & -30:18:30.46 & 15.47 & 9.53 & 440 & 81 & 69 & 247.00 $\pm$ 0.60\\
\hline
\end{tabular}
\tablefoot{$^{a}$ Total exposure time of an ABBA sequence.}
\end{table*}

\subsection{Target selection} \label{sec:target}
We first chose stars within 3$\arcmin$ and 2$\arcmin$ of the central region of Camargo~1103 ($l$ = 5.604$\degree$; $b$ = $-$2.121$\degree$) and Camargo~1106 ($l$ = 357.351$\degree$; $b$ = 1.683$\degree$). These are the regions in which \citet{Camargo2018} selected member stars. We used the 2MASS catalog \citep{Skrutskie2006} combined with Gaia DR2 \citep{GaiaCollaboration2018}.
In addition, we applied color and magnitude criteria for potential cluster member stars on the red giant branches in the ($K_{S}$, $J-K_{S}$) and ($G$, $BP-RP$) color-magnitude diagrams (CMDs) for Camargo~1103 (1.3 $<$ $J-K_{S}$ $<$ 2.2, 9 $<$ $K_{S}$ $<$ 12, 2.5 $<$ $BP-RP$ $<$ 5.0, and 15 $<$ $G$ $<$ 17.5) and Camargo~1106 (1.3 $<$ $J-K_{S}$ $<$ 2.0, 9 $<$ $K_{S}$ $<$ 12, 3.0 $<$ $BP-RP$ $<$ 5.0, and 15 $<$ $G$ $<$ 17). 
Finally, considering the position in the CMDs and contamination by adjacent stars, 15 and 10 target stars were selected from 72 and 87 member candidates for Camargo~1103 and 1106, respectively.
These stars also overlap in the proper motion plane with those of \citet{Camargo2018}. 
However, our procedure cannot guarantee that our target stars are identical to the stars selected by \citet{Camargo2018}
because \citet{Camargo2018} employed a statistically field-star decontaminated CMD for their cluster detection and CMD analysis.
We note that while Gaia DR2 was used for our target selection procedure, we adopt Gaia EDR3 \citep{GaiaCollaboration2021} information in the following analysis and figures.
Figure~\ref{fig:cmd} shows the coordinates, proper motions, and locations in the 2MASS and Gaia CMDs of the final target stars, and their basic information is given in Table~\ref{tab:info}. 
Although our targets in Camargo~1106 field are more widely scattered than those in Camargo~1103 field in the proper motion diagram, they still overlap with the proper motion criteria of \citet{Camargo2018}.
The stellar IDs are taken from the combination of the name of GC candidate and the last five digits of the 2MASS designation. 

\subsection{Observations} 
The observations using  IGRINS at the Gemini-South telescope were performed in service mode {over ten nights} between  February and May 2021 under program GS-2021A-Q-123 (PI: Sang-Hyun Chun). 
IGRINS is a cross-dispersed spectrograph with two separate arms covering the H- and K-bands. It provides a spectral coverage of 1.45$-$2.45 $\si{\micro\metre}$ at a spectral resolution of R$\sim$45,000 \citep{Park2014, Mace2018}.
Each  spectrum was taken in an ABBA nod sequence along the slit, together with a nearby A0V telluric standard star. 
The total exposure time of the ABBA sequence for each target is listed in Table~\ref{tab:info}. 

\subsection{Data reduction} 
The obtained IGRINS spectra were reduced with the IGRINS pipeline package \citep[PLP;][]{Lee2017}, which performs flat fielding, subtraction of the A and B images (A$-$B) to efficiently remove sky background, wavelength calibration using OH emission and telluric lines, and extraction of an optimal 1D spectrum based on the algorithm of \citet{Horne1986}. 
Because the IGRINS spectrum consists of 28 and 26 orders for the H and K arms, we combined these into a single continuous spectrum. 
The effective spectral ranges of each order are given in Table~\ref{tab:range}. 
After merging the orders, we performed continuum normalization using the {\em specutils} package of the {\em Astropy} \citep{AstropyCollaboration2013, AstropyCollaboration2018}. 
The wavelength scale of these spectra was then converted from vacuum into air using the formula of \citet{Morton2000}.

Radial velocities (RVs) of each star were determined by cross-correlation against a template spectrum obtained from the POLLUX database \citep{Palacios2010} using the {\em fxcor} task within the IRAF RV package. 
We separately estimated RVs from the H- and K-band spectra and then Doppler-corrected  each spectrum. 
We note that the difference in RV between H- and K-band spectra is smaller than 1.0 km~s$^{-1}$.
The heliocentric RVs (RV$_{helio}$) stated in Table~\ref{tab:info} are the straight means of the two values.  
We estimated signal-to-noise ratios (S/N) for H- and K-band spectra adopting the variances obtained  by the PLP during the  data reduction process. 
The S/N for H- and K-band spectra are listed separately in Table~\ref{tab:info}.

\begin{table}
\caption{Effective spectral ranges and echelle diffraction orders}
\label{tab:range} 
\centering                                    
\begin{tabular}{@{\extracolsep{3pt}} c c  c c}  
\hline\hline 
\multicolumn{2}{c}{H-band} & \multicolumn{2}{c}{K-band}  \\
\cline{1-2}  \cline{3-4} 
Order & Range ($\si{\micro\metre}$)     & Order & Range ($\si{\micro\metre}$) \\
\hline
99      & 1.797 -- 1.810        & 72 & 2.4506 -- 2.477 \\ 
100     & 1.780 -- 1.797        & 73 & 2.4175 -- 2.4506 \\ 
101     & 1.763 -- 1.780        & 74 & 2.3855 -- 2.4175 \\
102     & 1.746 -- 1.763        & 75 & 2.3545 -- 2.3855 \\
103     & 1.730 -- 1.746        & 76 & 2.324 -- 2.3545 \\
104     & 1.713 -- 1.730        & 77 & 2.295 -- 2.324 \\
105     & 1.698 -- 1.713        & 78 & 2.266 -- 2.295 \\
106     & 1.682 -- 1.698        & 79 & 2.238 -- 2.266 \\
107     & 1.667 -- 1.682        & 80 & 2.210 -- 2.238 \\
108     & 1.652 -- 1.667        & 81 & 2.184 -- 2.210 \\
109     & 1.638 -- 1.652        & 82 & 2.158 -- 2.184 \\
110     & 1.623 -- 1.638        & 83 & 2.133 -- 2.158 \\
111     & 1.610 -- 1.623        & 84 & 2.108 -- 2.133 \\
112     & 1.595 -- 1.610        & 85 & 2.084 -- 2.108 \\
113     & 1.582 -- 1.595        & 86 & 2.060 -- 2.084 \\
114     & 1.569 -- 1.582        & 87 & 2.037 -- 2.060 \\
115     & 1.556 -- 1.569        & 88 & 2.015 -- 2.037 \\
116     & 1.543 -- 1.556        & 89 & 1.993 -- 2.015 \\
117     & 1.530 -- 1.543        & 90 & 1.971 -- 1.993 \\
118     & 1.518 -- 1.530        & 91 & 1.950 -- 1.971 \\
119     & 1.505 -- 1.518        & 92 & 1.940 -- 1.950 \\
120     & 1.493 -- 1.505        & 93 & -- \\ 
121     & 1.482 -- 1.493        & 94 & -- \\
122     & 1.470 -- 1.482        & 95 & -- \\
123     & 1.457 -- 1.470        & 96 & -- \\
\hline                                             
\end{tabular}
\end{table}


\section{Spectroscopic analysis} \label{sec:analysis}
\subsection{Atmosphere parameters} \label{sec:param}
Deriving stellar parameters of bulge stars from photometry is challenging because of the high extinction and uncertain distances. 
The standard spectroscopic approach using EWs of Fe\,{\sc i} and Fe\,{\sc ii} lines is also restricted in NIR spectra by the absence of strong enough Fe\,{\sc ii} lines and strong contamination by molecular lines for more metal-rich systems. 

\subsubsection{Effective temperature} \label{sec:teff} 
As an example, we computed the effective temperature ($T_{\rm eff}$) from the various color-combinations of 2MASS and $Gaia$, such as $J-K_{S}$, $BP-RP$, and $G-K_{S}$, using the relations and coefficients of \citet{GonzalezHernandez2009} and \citet{Mucciarelli2020} 
with reddening corrections taken from the VVV-based map of \citet{Simion2017}, and from the Bayestar17, which provides 3D dust maps based on Pan-STARRS~1 and 2MASS photometry with Gaia parallaxes \citep{Green2018}.
For both maps, we used the extinction law of \citet{Green2018} with the coefficients of \citet{Nishiyama2009} and \citet{Casagrande2018} for the 2MASS and Gaia bands to derive the extinction values.
The average extinction values, E(B$-$V), are 1.79 and 1.73 mag for the Camargo~1103 field from the VVV and Bayestar17 maps, respectively. 
For Camargo~1106, an average E(B$-$V) of 1.58 mag is measured from the VVV, while no value is available in the Bayestar17 footprint. 
We note that much higher extinctions are given from the \citet{Schlegel1998} and \citet{Schlafly2011} maps, namely 2.54 versus 2.18 and 3.08 versus 2.61 mag for Camargo~1103 and 1106.
This leads to a typical difference of 300~K on the measured $T_{\rm eff}$ when the VVV or Bayestar17 extinction values are used in combination with the $J-K_{S}$ color, which is the color that is least affected by reddening. 
In addition, even when we adopt the same extinction value for each star, $T_{\rm eff}$ varies by $\sim$1000~K depending on the color index. 
For example, $T_{\rm eff}$ is estimated to be 4803~K from $J-K_{S}$, 5246~K from $BP-RP$, and 6209~K from $G-K_{S}$ for C1103-9023. 

\begin{figure*}
\centering
   \includegraphics[width=0.8\textwidth]{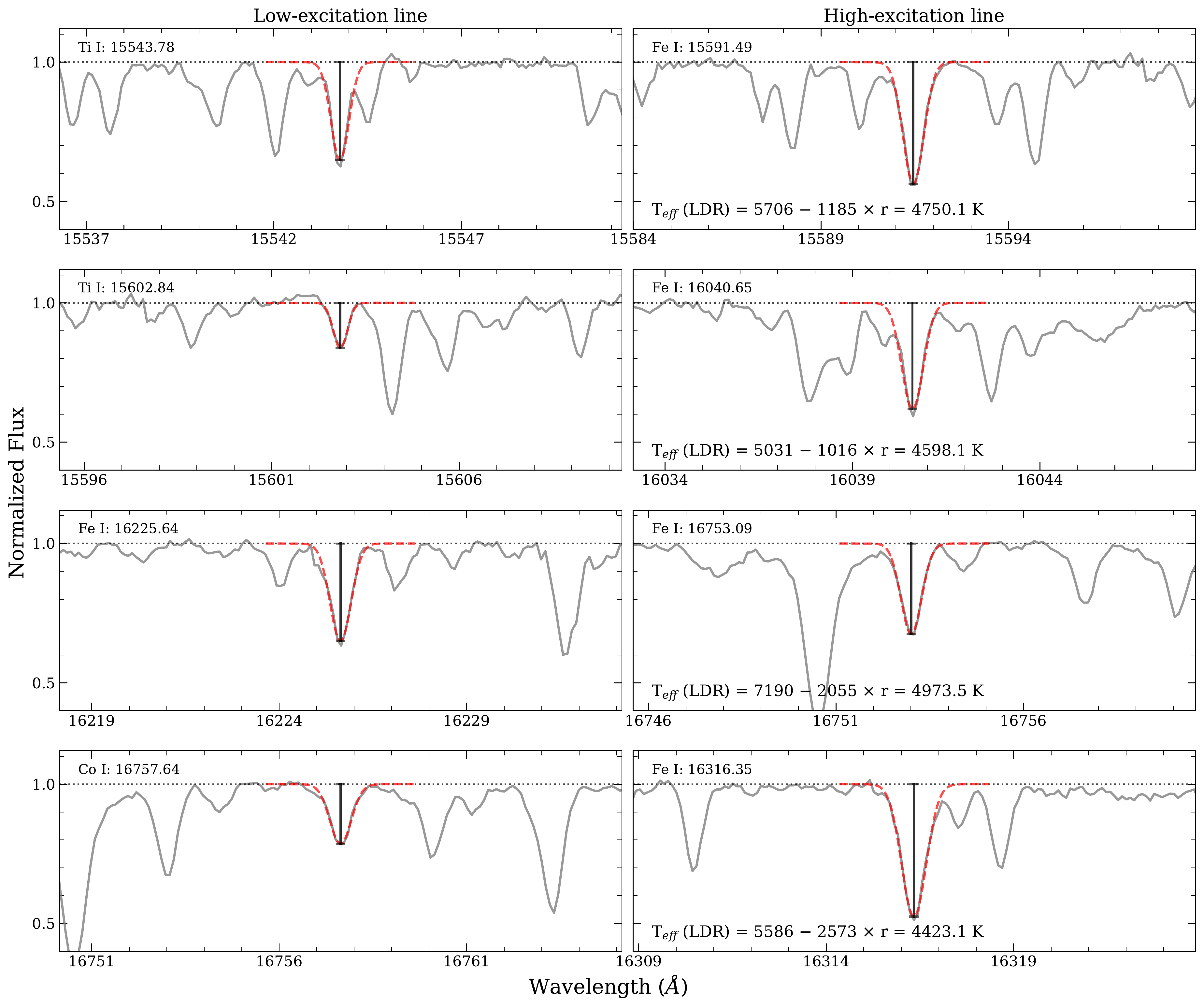}
     \caption{Examples of low-excitation and high-excitation line pairs used for the LDR method for C1103-10405 ($T_{\rm eff}$ = 4680~K). 
     The gray line is the observed spectrum, and the dashed red line is a Gaussian fit. 
     We measured the line depth as the maximum height of the fitted line measured from the continuum. 
     The LDR ($r$) is estimated as the ratio of the line depths of low- and high-excitation lines. 
     The relations between $T_{\rm eff}$ and $r$ for each pair of \citet{Fukue2015} are listed in the right panels.
     }
     \label{fig:ldr}
\end{figure*}

Instead of using the photometric approach, we therefore estimated $T_{\rm eff}$ from line-depth ratios (LDRs) following \citet{Fukue2015}, surface gravity ($\log{g}$) from the EW of the CO-overtone band \citep{Park2018}, and microturbulence velocity ($\xi_{t}$) from the relation between other parameters \citep{Mashonkina2017}.
The LDR method has often been used to derive $T_{\rm eff}$ by comparing the line depth of low- and high-excitation absorption lines \citep[see, e.g.,][]{Gray1991, Matsunaga2021}. 
In particular, \citet{Fukue2015} suggested nine-line pairs in the H-band region as $T_{\rm eff}$ indicators of giant stars. 
\citet{BocekTopcu2020} also confirmed that $T_{\rm eff}$ obtained from LDRs in IGRINS data shows a good agreement with that from optical spectroscopy.
We measured the LDR of these nine absorption pairs and estimated $T_{\rm eff}$ for each pair using the relations given in \citet{Fukue2015}.
The final $T_{\rm eff}$ for each target star was obtained as the mean, excluding the highest and lowest values. 
Figure~\ref{fig:ldr} shows example line pairs employed for the LDR method. 
Overall, we measured a given line depth by Gaussian profile fitting for each absorption line. 

\begin{figure}
\centering
   \includegraphics[width=0.35\textwidth]{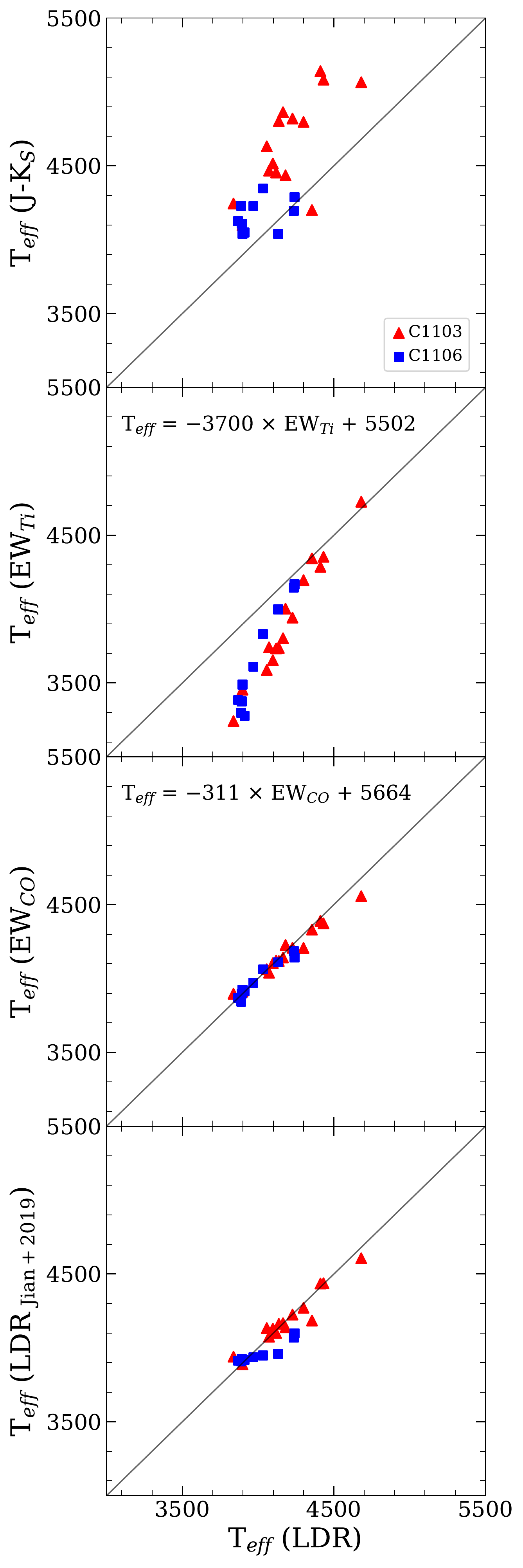}
     \caption{Comparisons of $T_{\rm eff}$ obtained from the LDR method with that from $J-K_{S}$ color, EW$_{Ti}$, EW$_{CO}$, and the other LDR method by \citet{Jian2019}.
     The red triangles and blue squares are target stars in the field of Camargo~1103 and 1106. 
     The $T_{\rm eff}$ values derived from the LDR method and EW$_{CO}$ are almost identical over the entire range.
     In addition, these values also agree well with $T_{\rm eff}$ derived from the relations of \citet{Jian2019}, which include metallicity terms.
     }
     \label{fig:teff}
\end{figure}

On the other hand, \citet{Park2018} suggested empirical relations to derive $T_{\rm eff}$ and $\log{g}$ from EWs of several lines in the NIR region.
We estimated $T_{\rm eff}$ from the Ti\,{\sc i} 2.224 $\si{\micro\metre}$ and CO 2.293 $\si{\micro\metre}$ features through these relations in order to compare the $T_{\rm eff}$ obtained from the LDR method.  
We note that the wavelengths of \citet{Park2018} are given in vacuum, and therefore we converted these into air.
Figure~\ref{fig:teff} shows comparisons of $T_{\rm eff}$ derived from the LDR method with those from $J-K_{S}$ colors, EW$_{Ti}$, EW$_{CO}$, and another LDR method of \citet{Jian2019} (see below).
$T_{\rm eff}$ from $J-K_{S}$ color in concert with reddening from the VVV reddening agrees well with the LDR-based $T_{\rm eff}$ for stars in Camargo~1106 field with a mean difference of 185$\pm$108~K (1$\sigma$ scatter), while there is an offset of 465$\pm$184~K for stars in the Camargo~1103 field. 
This is probably due to the fact that the extinction and its uncertainty in the Camargo~1103 field are more severe than those in Camargo~1106.  
In the case of $T_{\rm eff}$ estimated from EW$_{Ti}$ (the second panel of Figure~\ref{fig:teff}), these values correlate with the LDR-based $T_{\rm eff}$, but they are not identical.
In particular, the discrepancy between the two measurements increases with decreasing $T_{\rm eff}$. 

Interestingly, the temperatures obtained from the LDR method and EW of CO line are almost identical, with a mean difference of 33$\pm$32~K.
The differences between the two estimates are smaller than 10~K for seven stars, and even the largest difference is only 120~K (C1103-10405).
It is comforting that the two estimates, independently measured from the H- (LDR) and K-band (EW$_{CO}$) spectra, agree well.
The very similar results of the two independent estimates may appear surprising. 
We note that \citet{Fukue2015} used ten solar neighborhood stars to derive LDR-$T_{\rm eff}$ relations, whereas \citet{Park2018} obtained EW$_{CO}$-$T_{\rm eff}$ relation from 48 MK standard stars without common samples between the two studies. 
The similar values obtained from the two separate methods therefore mean that our estimates are close to the actual $T_{\rm eff}$ of stars.
We adopted $T_{\rm eff}$ from the LDRs as the final value from these two values because this value is based on more absorption lines, while the other is derived from a single CO-feature. 
The errors on the parameter determinations are discussed in detail in Section~\ref{sec:err}.

On the other hand, \citet{Jian2019} reported that the LDR-based $T_{\rm eff}$ depends on stellar metallicity.
They suggested 11-line pairs (7 pairs are in common with \citealt{Fukue2015}) with individual relations including [Fe/H] and abundance ratio terms.
We estimated $T_{\rm eff}$ using the relations given in \citet{Jian2019} after abundance measurements and compared them with those adopted in this study.
In the same manner as our LDR-based $T_{\rm eff}$ estimate, we excluded the highest and lowest values from the 11 pairs to derive the mean value.
The eighth line pair was also excluded because of the large uncertainty in the Co abundance measurement.
As shown in the bottom panel of Figure~\ref{fig:teff}, these two estimates agree well with a mean difference of 53$\pm$54~K.
While the differences are increased for metal-poor stars, the largest difference is still 172~K for C1106-15096 with [Fe/H] = $-$0.56~dex.
This comparison underlines that our $T_{\rm eff}$ determination is reliable, although we did not use the $T_{\rm eff}$ derived from the \citet{Jian2019} method in the following analysis.

\subsubsection{Surface gravity}
\citet{Park2018} reported that the EW of the CO 2.293 $\si{\micro\metre}$ line correlates not only with $T_{\rm eff}$ , but also with $\log{g}$ as follows:
\begin{equation} \label{eq:logg}
\begin{aligned}
\log{g} =       & ~(-2.130 \pm 0.003)~{\rm EW}_{CO} + (-41.79 \pm 0.05)~\log{T_{\rm eff}} \\
                & + (162.54 \pm 0.18). 
\end{aligned}
\end{equation}
We estimated $\log{g}$ for our target stars using this relation. 
We note again that it is difficult to derive $\log{g}$ from the ionization equilibrium between Fe\,{\sc i} and Fe\,{\sc ii} or from the canonical relation using the bolometric magnitude because only a small number of Fe\,{\sc ii} absorptions is available in the NIR spectrum and the magnitudes and distances of stars toward the bulge are very inaccurate. 
The derived $\log{g}$ values are mostly distributed in the range of 1.0 to 2.0 dex, which are realistic values for bright giant stars, {as supported by the Kiel diagrams in Figure~\ref{fig:apo_param}.}

\subsubsection{Microturbulence}
For the microturbulence, $\xi_{t}$, we used the relation of \citet{Boeche2016}, which reads as follows:
\begin{equation} \label{eq:vt1}
\xi_{t}~(BC16) = \sum_{i,j=0}^{2} a_{ij}(T_{\rm eff})^{i}(\log{g})^{j}, 
\end{equation}
where $a_{00}$ = $-$5.11308, $a_{01}$ = 0.58507, $a_{02}$ = 0.471885,
$a_{10}$ = 0.00207105, $a_{11}$ = 1.70456 $\times$ 10$^{-5}$, $a_{12}$ = $-$0.000257162
$a_{20}$ = $-$1.0543 $\times$ 10$^{-7}$, $a_{21}$ = $-$3.21628 $\times$ 10$^{-8}$, 
and $a_{22}$ = 2.94647 $\times$ 10$^{-8}$.
This was compared to the following relation as taken from  \citet{Mashonkina2017}:
\begin{equation} \label{eq:vt2}
\xi_{t}~(M17) = 0.14 - 0.08 \times {\rm [Fe/H]} + 4.90 \times (T_{\rm eff}/10^{4}) - 0.47 \times \log{g}. 
\end{equation}
Because Eq~(\ref{eq:vt2}) requires a term of [Fe/H], we first estimated $\xi_{t}$ using Eq~(\ref{eq:vt1}), which has no metallicity dependence. 
We measured temporary [Fe/H] through spectrum synthesis with this $\xi_{t}~(BC16)$ and $T_{\rm eff}$ and $\log{g}$ as derived above. 
Then, we reestimated $\xi_{t}$ using Eq~(\ref{eq:vt2}). 
The final $\xi_{t}~(M17)$ and [Fe/H] values of the model atmospheres were obtained from the iteration of the [Fe/H] measurement until the newly estimated value was the same as the input.
Although these two estimates of $\xi_{t}$ show a small difference ($\sim$ 0.16 km~s$^{-1}$), we used $\xi_{t}~(M17)$ as the final value because they agree better with the trend of APOGEE parameters (see Figure~\ref{fig:apo_param}).
The spectroscopic method for deriving $\xi_{t}$, which removes the trend between abundance and reduced EW, could not be applied, because the EWs of Fe-lines in the NIR spectra are highly contaminated by molecular lines.
The final atmosphere parameters for each star are listed in Table~\ref{tab:param}.

\begin{table}
\caption{Atmosphere parameters}
\label{tab:param} 
\centering                                    
\begin{tabular}{l c c c c}  
\hline\hline 
\multirow{2}{*}{ID}     & $T_{\rm eff}$ & $\log{g}$     & $\xi_{t}$                     & [Fe/H]$_{model}$ \\ 
                                & [K]                   & [dex]         & [km~s$^{-1}$]   & [dex] \\
\hline                           
C1103-7052 & 3897 & 1.16 & 1.52 & $-$0.17 \\
C1103-8510 & 4072 & 1.24 & 1.56 & $-$0.08 \\
C1103-9023 & 4137 & 1.50 & 1.46 & 0.06 \\
C1103-9100 & 4355 & 2.02 & 1.37 & $-$0.59 \\
C1103-9180 & 4411 & 2.19 & 1.25 & 0.25 \\
C1103-9181 & 3837 & 1.34 & 1.38 & 0.16 \\
C1103-9341 & 4181 & 2.05 & 1.21 & 0.19 \\
C1103-9590 & 4227 & 1.72 & 1.39 & 0.12 \\
C1103-10048 & 4119 & 1.60 & 1.39 & 0.16 \\
C1103-10405 & 4680 & 2.26 & 1.35 & 0.21 \\
C1103-10484 & 4164 & 1.54 & 1.44 & 0.16 \\
C1103-11170 & 4299 & 1.41 & 1.59 & $-$0.09 \\
C1103-11530 & 4431 & 2.00 & 1.35 & 0.21 \\
C1103-12006 & 4097 & 1.58 & 1.39 & 0.16 \\
C1103-12308 & 4057 & 1.48 & 1.42 & 0.15 \\
\hline
C1106-15096 & 4132 & 1.46 & 1.52 & $-$0.56 \\
C1106-15592 & 3867 & 1.01 & 1.57 & $-$0.07 \\
C1106-16402 & 3897 & 1.25 & 1.48 & $-$0.21 \\
C1106-16451 & 3911 & 1.11 & 1.53 & 0.12 \\
C1106-16566 & 3968 & 1.24 & 1.52 & $-$0.22 \\
C1106-17160 & 4235 & 1.52 & 1.55 & $-$0.62 \\
C1106-17565 & 3888 & 0.73 & 1.71 & $-$0.11 \\
C1106-18086 & 4241 & 1.21 & 1.70 & $-$0.69 \\
C1106-18101 & 4032 & 1.58 & 1.41 & $-$0.44 \\
C1106-18303 & 3891 & 1.09 & 1.54 & $-$0.05 \\
\hline
\end{tabular}
\tablefoot{[Fe/H]$_{model}$ indicates the input value for the atmosphere model.}
\end{table}
%
\subsubsection{Comparison with APOGEE}
\begin{figure}
\centering
   \includegraphics[width=0.48\textwidth]{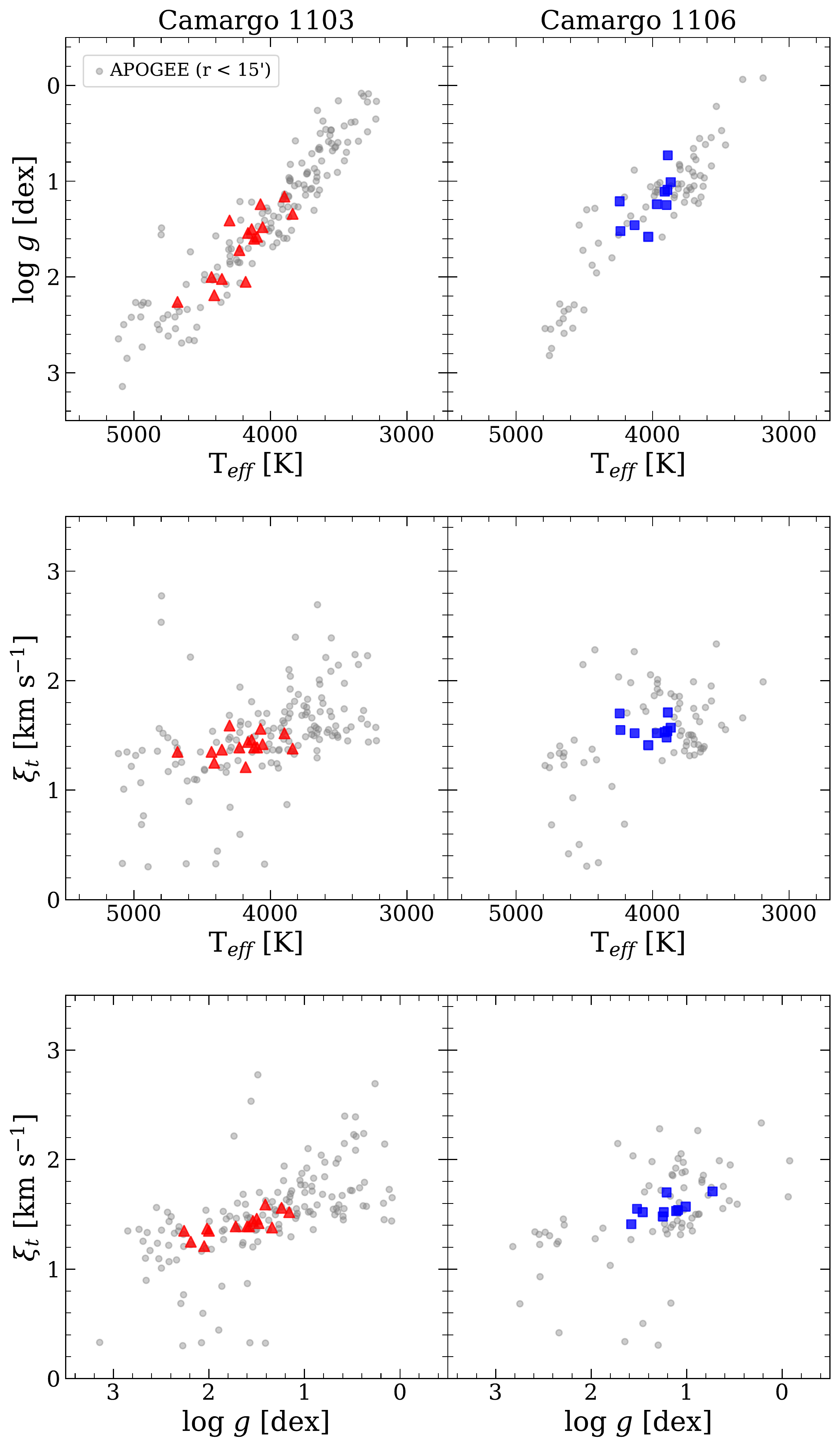}
     \caption{Atmospheric parameters of our target stars, together with APOGEE sample stars within 15$'$ radius around Camargo~1103 and 1106.
     The symbols are the same as in Figure~\ref{fig:cmd}.
     Our derived parameters follow the trends of APOGEE data on the $T_{\rm eff} - \log{g}$, $T_{\rm eff} - \xi_{t}$, and $\log{g} - \xi_{t}$ planes well.
     }
     \label{fig:apo_param}
\end{figure}

In order to confirm the validity of our atmospheric parameters, we compared them  with those of nearby stars ($<$ 15$\arcmin$) from the APOGEE DR16 \citep{Ahumada2020}, which are determined by the entire spectrum fitting \citep{Jonsson2020}. 
As shown in Figure~\ref{fig:cmd}, our target stars are placed in the color and magnitude range of the APOGEE stars.
Figure~\ref{fig:apo_param} shows the comparison with APOGEE parameters on the $T_{\rm eff} - \log{g}$, $T_{\rm eff} - \xi_{t}$, and $\log{g} - \xi_{t}$ planes. 
All these plots demonstrate that our atmospheric parameters well follow the local trends of the APOGEE stars. 
These comparisons, therefore, support that our parameter estimates are reliable.  
We note that only one of our stars (C1106-18101) is in common with the local APOGEE sample. 
The derived $T_{\rm eff}$, $\log{g}$, $\xi_{t}$, and [Fe/H] are 4032~K, 1.58~dex, 1.41~km~s$^{-1}$, and $-$0.44~dex from our study, and they are 3865~K, 1.05~dex, 1.88~km~s$^{-1}$, and $-$0.54~dex from the APOGEE.

On the other hand, if we estimated $\log{g}$ employing the $T_{\rm eff}$ from EW$_{Ti}$, these parameters would be very different from the APOGEE parameters.
The $\log{g}$ values increase with decreasing temperature, and therefore stars with $T_{\rm eff}$ $<$ 3500~K have $\log{g}$ $>$ 4.0~dex, unlike the general trend of giant stars.  
Therefore, we conclude as above that $T_{\rm eff}$ derived from the LDR method or EW$_{CO}$ is more reliable than that from EW$_{Ti}$ for our target stars. 

\subsection{Chemical abundance determination} \label{sec:abund}
With the determined atmosphere parameters, we first generated model atmospheres for each star using both the ATLAS9 grid of Kurucz \citep{Castelli2003} and a grid of spherical MARCS models \citep{Gustafsson2008}.  
The chemical abundances were measured using the 2019NOV version of the local thermodynamic equilibrium (LTE) code MOOG \citep{Sneden1973}. 
We performed spectrum synthesis to measure the abundances of Fe, Na, Mg, Al, Si, S, K, Ca, Ti, Cr, Ni, and Ce using MOOG's {\em synth} driver. 
We employed the line list of \citet{Afsar2018}, which has been specifically generated for IGRINS spectroscopy. 
However, several lines were excluded because they have a low S/N or because the absorption lines are severely blended (e.g., Sc and Co lines). 
The line information is presented in Table~\ref{tab:line}. 
In addition to this line list, we used the Kurucz line list\footnote{http://kurucz.harvard.edu/linelists.html}, including CN, CO, and C$_{2}$ features, to generate a synthetic spectrum for the full range. 
We measured abundances for each line through a comparison with synthetic spectra for the central $\pm$0.2~$\AA$ $\sim$ $\pm$0.8~$\AA$ spectral region from the peak of the absorption to minimize the blending contamination. 
The synthetic spectra were broadened to fit the observed spectra using the Gaussian smoothing function of MOOG. 
We note that the actual spectral resolution of IGRINS data varies depending on the position of each order from 38000 to 45000.
We performed a visual inspection of each line to find the best-fit synthetic model by changing the fitting range, continuum level, and smoothing parameter.
The abundance was then measured where the residual, the squared sum of discrepancy of the relative flux between the observed and synthetic spectrum, was minimum in the central spectral region.
The final abundances for each element were derived as the mean of the measurements for each line.

\begin{figure}
\centering
   \includegraphics[width=0.32\textwidth]{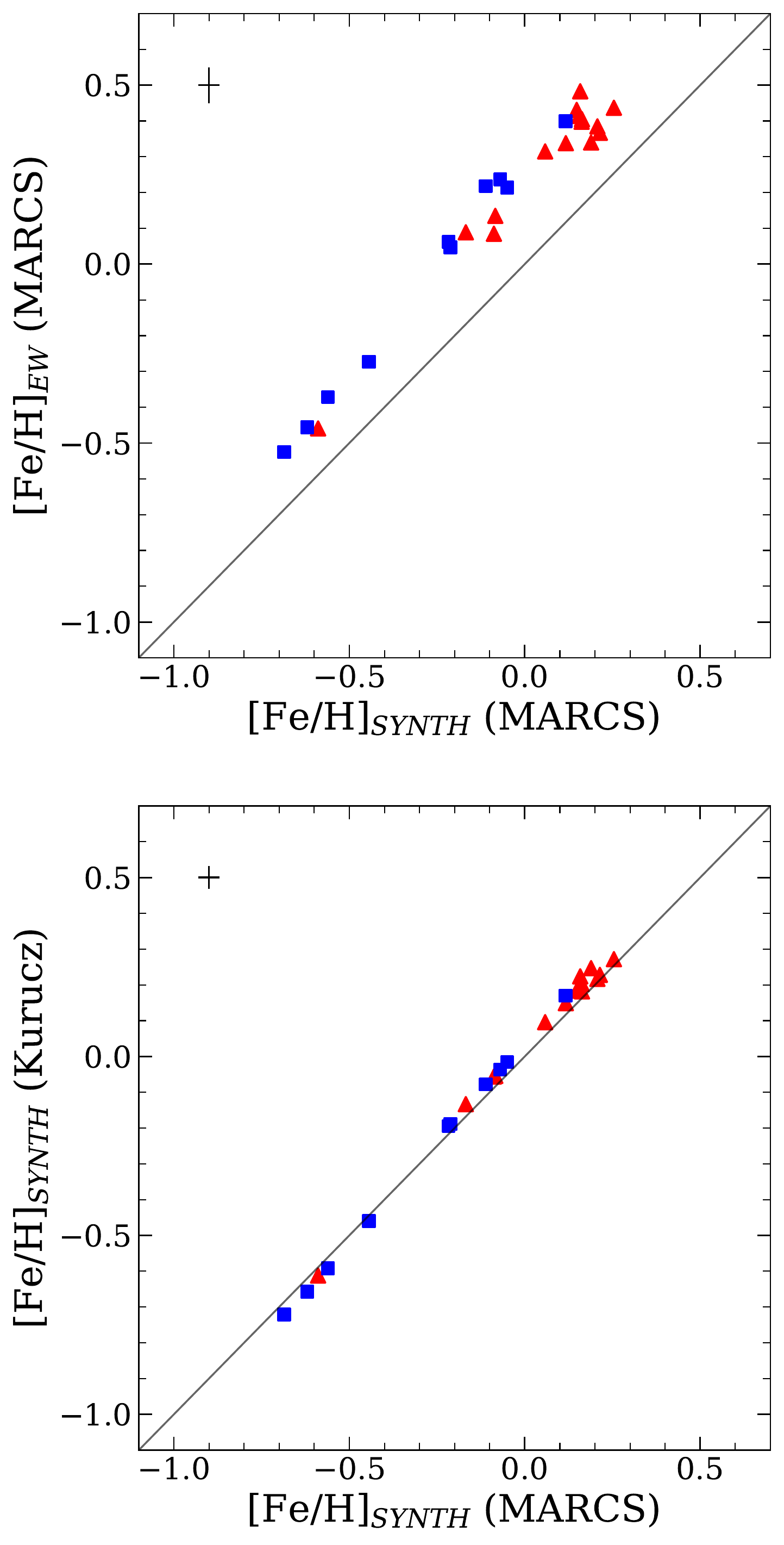}
     \caption{Comparison of [Fe/H] abundance ratio derived from spectrum synthesis and EW measurement with MARCS models (upper panel), and comparison of [Fe/H] using MARCS and Kurucz models (lower panel).
     The symbols are the same as in Figure~\ref{fig:teff}.
     The typical error of each measurement is plotted in the upper left corner.
     }
     \label{fig:fe}
\end{figure}

It is important to note that in the chemical abundance measurement on the NIR spectrum, spectral synthesis is preferred over EW measurements because of substantial contamination by molecular lines.  
For instance, in the upper panel of Figure~\ref{fig:fe}, we compare the [Fe/H] abundance ratios derived from EWs and spectrum synthesis on identical Fe absorption lines, which are very likely not affected by blending from the comparison with synthetic spectra. 
The EWs were measured with a Gaussian fitting for each absorption line using the {\em specutils} package of {\em Astropy}.
The [Fe/H] values from EWs are generally higher than those from spectrum synthesis, with a mean difference of 0.23~dex. 
The typical line-to-line scatter on the Fe-abundance measurement from EWs is also larger than that from spectral synthesis (0.25~dex versus 0.15~dex).
\citet{Afsar2018} reported, however, that the Fe abundances derived from EW measurement and spectrum synthesis agreed well ($<$ 0.1 dex). 
The two studies might differ because the S/N of the sample used by \citet{Afsar2018} may have been higher or because the metallicity of our target stars is higher. 

In the lower panel of Figure~\ref{fig:fe}, we also compare the [Fe/H] abundance ratios derived from spectrum synthesis employing two different atmosphere models, Kurucz and MARCS.
These values are almost identical, with a typical difference of 0.03~dex.
We finally adopted the MARCS models for our spectroscopic analysis because the line-to-line variation using this model is slightly weaker than the variation we obtained with the  Kurucz model. 

\begin{table}
\small
\setlength{\tabcolsep}{1pt}
\caption{Line list}
\label{tab:line} 
\centering                                    
\begin{tabular}{@{\extracolsep{1pt}} l c c c @{\hskip 5pt}|@{\hskip 5pt} l c c c }   
\hline\hline 
$\lambda$       & Species       & EP            & $\log{gf}$    & $\lambda$     & Species & EP            & $\log{gf}$  \\
\hline
$[\AA]$         &               & [eV]  &                       & $[\AA]$               &               & [eV]    & \\  
\hline
22056.426 & Na\,{\sc i} & 3.19 & 0.29           & 15602.842$^{*}$ & Ti\,{\sc i} & 2.27 & $-$1.59 \\
22083.661 & Na\,{\sc i} & 3.19 & $-$0.01      & 17376.577 & Ti\,{\sc i} & 4.49 & 0.55 \\
23348.423 & Na\,{\sc i} & 3.75 & 0.28           & 17383.103 & Ti\,{\sc i} & 4.48 & 0.44 \\
23379.136 & Na\,{\sc i} & 3.75 & 0.54           & 21782.944$^{K,*}$ & Ti\,{\sc i} & 1.75 & $-$1.14 \\
15024.997$^{*}$ & Mg\,{\sc i} & 5.11 & 0.36     & 21897.376$^{K,*}$ & Ti\,{\sc i} & 1.74 & $-$1.44 \\
15040.246$^{*}$ & Mg\,{\sc i} & 5.10 & 0.14     & 22004.500$^{K}$ & Ti\,{\sc i} & 1.73 & $-$1.85 \\
15047.714$^{*}$ & Mg\,{\sc i} & 5.11 & $-$0.34    & 22211.238$^{K,*}$ & Ti\,{\sc i} & 1.73 & $-$1.75 \\
15740.705$^{*}$ & Mg\,{\sc i} & 5.93 & $-$0.21    & 22232.858$^{K,*}$ & Ti\,{\sc i} & 1.74 & $-$1.62 \\
15748.988$^{*}$ & Mg\,{\sc i} & 5.93 & 0.14     & 22443.925$^{K,*}$ & Ti\,{\sc i} & 1.74 & $-$2.30 \\
15765.839$^{*}$ & Mg\,{\sc i} & 5.93 & 0.41     & 15680.060$^{*}$ & Cr\,{\sc i} & 4.70 & 0.15 \\
17108.631$^{*}$ & Mg\,{\sc i} & 5.39 & 0.06     & 15860.210$^{*}$ & Cr\,{\sc i} & 4.70 & 0.00 \\
21213.725$^{*}$ & Mg\,{\sc i} & 6.73 & $-$1.60    & 17708.730$^{*}$ & Cr\,{\sc i} & 4.39 & $-$0.51 \\
21225.620$^{*}$ & Mg\,{\sc i} & 6.73 & $-$1.38    & 15194.490$^{*}$ & Fe\,{\sc i} & 2.22 & $-$4.75 \\
21458.865$^{*}$ & Mg\,{\sc i} & 6.52 & $-$1.32    & 15207.526$^{*}$ & Fe\,{\sc i} & 5.39 & 0.08 \\
16763.369 & Al\,{\sc i} & 4.09 & $-$0.48          & 15343.788$^{*}$ & Fe\,{\sc i} & 5.65 & $-$0.69 \\
17699.050 & Al\,{\sc i} & 4.67 & $-$1.21          & 15493.515$^{*}$ & Fe\,{\sc i} & 6.36 & $-$1.06 \\
21093.078$^{K}$ & Al\,{\sc i} & 4.09 & $-$0.40          & 15648.510$^{*}$ & Fe\,{\sc i} & 5.43 & $-$0.70 \\
21163.800$^{K}$ & Al\,{\sc i} & 4.09 & $-$0.09          & 15662.013 & Fe\,{\sc i} & 5.83 & 0.07 \\
21208.176$^{K}$ & Al\,{\sc i} & 5.12 & $-$0.46          & 15761.313 & Fe\,{\sc i} & 6.25 & $-$0.16 \\
22700.914$^{K}$ & Al\,{\sc i} & 4.83 & $-$1.49          & 15858.657$^{*}$ & Fe\,{\sc i} & 5.58 & $-$1.25 \\
15960.080$^{*}$ & Si\,{\sc i} & 5.98 & 0.20     & 15980.725$^{*}$ & Fe\,{\sc i} & 6.26 & 0.72 \\
16060.021$^{*}$ & Si\,{\sc i} & 5.95 & $-$0.50    & 16009.610$^{*}$ & Fe\,{\sc i} & 5.43 & $-$0.55 \\
16094.797$^{*}$ & Si\,{\sc i} & 5.96 & $-$0.09    & 16153.247$^{*}$ & Fe\,{\sc i} & 5.35 & $-$0.73 \\
16163.714$^{*}$ & Si\,{\sc i} & 5.95 & $-$0.95    & 16165.029 & Fe\,{\sc i} & 6.32 & 0.75 \\
16215.691$^{*}$ & Si\,{\sc i} & 5.95 & $-$0.58    & 16171.930 & Fe\,{\sc i} & 6.38 & $-$0.51 \\
16434.929$^{*}$ & Si\,{\sc i} & 5.96 & $-$1.49    & 17420.825$^{*}$ & Fe\,{\sc i} & 3.88 & $-$3.52 \\
16680.770$^{*}$ & Si\,{\sc i} & 5.98 & $-$0.09    & 21178.155$^{*}$ & Fe\,{\sc i} & 3.02 & $-$4.24 \\
19928.919$^{*}$ & Si\,{\sc i} & 6.10 & $-$0.33    & 21238.466$^{*}$ & Fe\,{\sc i} & 4.96 & $-$1.37 \\
20343.887$^{*}$ & Si\,{\sc i} & 6.13 & $-$1.13    & 21284.348 & Fe\,{\sc i} & 3.07 & $-$4.51 \\
22537.686$^{*}$ & Si\,{\sc i} & 6.62 & $-$0.30    & 21735.457$^{*}$ & Fe\,{\sc i} & 6.18 & $-$0.73 \\
22665.777$^{*}$ & Si\,{\sc i} & 6.62 & $-$0.47    & 21851.381$^{*}$ & Fe\,{\sc i} & 3.64 & $-$3.63 \\
15403.790 & S\,{\sc i} & 8.70 & 0.40            & 22257.107$^{*}$ & Fe\,{\sc i} & 5.06 & $-$0.82 \\
15469.816 & S\,{\sc i} & 8.05 & $-$0.26           & 22260.179$^{*}$ & Fe\,{\sc i} & 5.09 & $-$0.98 \\
22507.597 & S\,{\sc i} & 7.87 & $-$0.48           & 22385.102$^{*}$ & Fe\,{\sc i} & 5.32 & $-$1.57 \\
22519.106 & S\,{\sc i} & 7.87 & $-$0.38           & 22392.878$^{*}$ & Fe\,{\sc i} & 5.10 & $-$1.32 \\
22526.052 & S\,{\sc i} & 7.87 & $-$0.70           & 22419.976$^{*}$ & Fe\,{\sc i} & 6.22 & $-$0.30 \\
22563.868 & S\,{\sc i} & 7.87 & $-$0.28           & 22473.263 & Fe\,{\sc i} & 6.12 & 0.32 \\
22575.434 & S\,{\sc i} & 7.87 & $-$0.80           & 22619.838$^{*}$ & Fe\,{\sc i} & 4.99 & $-$0.51 \\
22707.736 & S\,{\sc i} & 7.87 & 0.18            & 23308.477$^{*}$ & Fe\,{\sc i} & 4.08 & $-$2.73 \\
15163.090 & K\,{\sc i} & 2.67 & 0.55            & 16310.501 & Ni\,{\sc i} & 5.28 & $-$0.02 \\
16136.823 & Ca\,{\sc i} & 4.53 & $-$0.67          & 16815.472 & Ni\,{\sc i} & 5.31 & $-$0.59 \\
16150.762 & Ca\,{\sc i} & 4.53 & $-$0.28          & 16818.745 & Ni\,{\sc i} & 6.04 & 0.33 \\
16155.236 & Ca\,{\sc i} & 4.53 & $-$0.77          & 16867.283 & Ni\,{\sc i} & 5.47 & $-$0.01 \\
16157.364 & Ca\,{\sc i} & 4.55 & $-$0.24          & 17306.518 & Ni\,{\sc i} & 5.49 & $-$0.53 \\
19815.017$^{K}$ & Ca\,{\sc i} & 4.62 & 0.40           & 15277.650 & Ce\,{\sc ii} & 0.61 & $-$1.94 \\
19933.727$^{K}$ & Ca\,{\sc i} & 3.91 & 0.16           & 15784.750 & Ce\,{\sc ii} & 0.32 & $-$1.54 \\
22607.944$^{K}$ & Ca\,{\sc i} & 4.68 & 0.43           & 15829.830 & Ce\,{\sc ii} & 0.32 & $-$1.80 \\
22624.962$^{K}$ & Ca\,{\sc i} & 4.68 & 0.62           & 16376.480 & Ce\,{\sc ii} & 0.12 & $-$1.79 \\
22626.723$^{K}$ & Ca\,{\sc i} & 4.68 & $-$0.32          & 16595.230 & Ce\,{\sc ii} & 0.12 & $-$2.19 \\
\hline
\end{tabular}
\tablefoot{The $\text{asterisk}$ indicates a line for which NLTE correction is available.
The superscript $K$ denotes a line that was excluded from the analysis because of the abundance difference between H- and K-band spectra (see Section~\ref{sec:HK}).}
\end{table}

\begin{table*}
\tiny
\setlength{\tabcolsep}{2.0pt}
\caption{Chemical abundance results for Fe, Na, Mg, Al, and Si.}
\label{tab:abund} 
\centering
\begin{tabular}{@{\extracolsep{2pt}}c r r r  r r r  r r r  r r r  r r r  r r r  r r r  r r r }   
\hline\hline 
\multirow{2}{*}{ID}             & \multicolumn{3}{c}{Fe}        & \multicolumn{3}{c}{Fe$_{\rm NLTE}$} & \multicolumn{3}{c}{Na}        & \multicolumn{3}{c}{Mg}        
                                & \multicolumn{3}{c}{Mg$_{\rm NLTE}$}   & \multicolumn{3}{c}{Al$_{\rm H-band}$}   & \multicolumn{3}{c}{Si}        & \multicolumn{3}{c}{Si$_{\rm NLTE}$} \\
\cline{2-4} \cline{5-7} \cline{8-10} \cline{11-13} \cline{14-16} \cline{17-19} \cline{20-22} \cline{23-25}
                                        & [X/H] & $\sigma$ & $N$        & [X/H] & $\sigma$ & $N$ & [X/H] & $\sigma$ & $N$ & [X/H] & $\sigma$ & $N$ 
                                        & [X/H] & $\sigma$ & $N$ & [X/H] & $\sigma$ & $N$ & [X/H] & $\sigma$ & $N$ & [X/H] & $\sigma$ & $N$ \\
\hline   
C1103-7052 & $-$0.17 &  0.15 & 26 & $-$0.17 &  0.16 & 20 &  0.29 &  0.07 &  3 &  0.02 &  0.13 &  8 & $-$0.09 &  0.15 &  8 &  0.18 &  0.02 &  2 & $-$0.29 &  0.17 &  8 & $-$0.34 &  0.18 &  8 \\
C1103-8510 & $-$0.08 &  0.17 & 26 & $-$0.12 &  0.18 & 20 &  0.55 &  0.03 &  3 &  0.13 &  0.13 &  8 &  0.00 &  0.17 &  8 &  0.14 &  0.09 &  2 & $-$0.11 &  0.17 &  8 & $-$0.18 &  0.19 &  8 \\
C1103-9023 &  0.06 &  0.15 & 25 &  0.02 &  0.14 & 19 &  0.77 &  0.07 &  3 &  0.23 &  0.10 &  8 &  0.11 &  0.10 &  8 &  0.26 &  0.10 &  2 &  0.10 &  0.18 &  8 &  0.03 &  0.17 &  8 \\
C1103-9100 & $-$0.59 &  0.14 & 26 & $-$0.58 &  0.13 & 21 & $-$0.55 &  0.10 &  2 & $-$0.35 &  0.15 &  8 & $-$0.46 &  0.19 &  8 & $-$0.29 &  0.14 &  2 & $-$0.36 &  0.20 & 11 & $-$0.43 &  0.21 & 11 \\
C1103-9180 &  0.25 &  0.14 & 25 &  0.21 &  0.13 & 19 &  0.68 &  0.03 &  3 &  0.36 &  0.13 &  8 &  0.29 &  0.14 &  8 &  0.33 &  0.14 &  2 &  0.32 &  0.14 &  8 &  0.25 &  0.14 &  8 \\
C1103-9181 &  0.16 &  0.19 & 25 &  0.14 &  0.20 & 19 &  0.94 &  0.07 &  3 &  0.19 &  0.11 &  9 &  0.10 &  0.12 &  7 &  0.32 &  0.14 &  2 &  0.06 &  0.27 &  6 &  0.01 &  0.28 &  6 \\
C1103-9341 &  0.19 &  0.16 & 26 &  0.20 &  0.16 & 20 &  0.28 &  0.02 &  3 &  0.09 &  0.14 &  9 &  0.01 &  0.15 &  9 &  0.14 &  0.10 &  2 &  0.06 &  0.14 &  8 &  0.00 &  0.15 &  8 \\
C1103-9590 &  0.12 &  0.16 & 26 &  0.10 &  0.15 & 20 &  0.76 &  0.06 &  3 &  0.22 &  0.12 & 10 &  0.12 &  0.14 &  8 &  0.26 &  0.09 &  2 &  0.08 &  0.20 &  8 &  0.01 &  0.21 &  8 \\
C1103-10048 &  0.16 &  0.17 & 26 &  0.14 &  0.18 & 20 &  0.75 &  0.10 &  4 &  0.24 &  0.15 &  9 &  0.14 &  0.15 &  9 &  0.20 &  0.15 &  2 &  0.03 &  0.15 &  8 & $-$0.04 &  0.15 &  8 \\
C1103-10405 &  0.21 &  0.16 & 26 &  0.20 &  0.16 & 20 &  0.63 &  0.03 &  3 &  0.26 &  0.11 & 10 &  0.18 &  0.10 & 10 &  0.20 &  0.10 &  2 &  0.20 &  0.17 & 10 &  0.12 &  0.18 & 10 \\
C1103-10484 &  0.16 &  0.18 & 25 &  0.13 &  0.19 & 19 &  0.75 &  0.04 &  4 &  0.30 &  0.15 &  9 &  0.20 &  0.16 &  9 &  0.15 &  0.15 &  2 &  0.15 &  0.15 &  8 &  0.07 &  0.16 &  8 \\
C1103-11170 & $-$0.09 &  0.13 & 26 & $-$0.11 &  0.14 & 20 &  0.35 &  0.07 &  3 &  0.11 &  0.10 &  9 & $-$0.00 &  0.11 &  9 &  0.04 &  0.12 &  2 & $-$0.17 &  0.08 &  7 & $-$0.25 &  0.10 &  7 \\
C1103-11530 &  0.21 &  0.16 & 25 &  0.17 &  0.14 & 20 &  0.78 &  0.02 &  3 &  0.33 &  0.12 &  8 &  0.25 &  0.12 &  8 &  0.23 &  0.10 &  2 &  0.27 &  0.18 &  9 &  0.19 &  0.18 &  9 \\
C1103-12006 &  0.16 &  0.18 & 24 &  0.15 &  0.18 & 18 &  0.84 &  0.04 &  4 &  0.26 &  0.17 &  7 &  0.16 &  0.17 &  7 &  0.22 &  0.10 &  2 &  0.10 &  0.20 &  7 &  0.04 &  0.20 &  7 \\
C1103-12308 &  0.15 &  0.17 & 25 &  0.11 &  0.17 & 19 &  0.96 &  0.06 &  3 &  0.29 &  0.10 &  8 &  0.19 &  0.11 &  8 &  0.28 &  0.07 &  2 &  0.09 &  0.21 &  9 &  0.02 &  0.21 &  9 \\
\hline
C1106-15096 & $-$0.56 &  0.14 & 25 & $-$0.57 &  0.15 & 20 & $-$0.49 &  0.09 &  3 & $-$0.37 &  0.15 &  8 & $-$0.50 &  0.17 &  8 & $-$0.49 &  0.12 &  2 & $-$0.46 &  0.13 & 10 & $-$0.52 &  0.15 & 10 \\
C1106-15592 & $-$0.07 &  0.16 & 26 & $-$0.09 &  0.17 & 20 &  0.34 &  0.19 &  2 &  0.07 &  0.18 &  8 & $-$0.05 &  0.19 &  8 & $-$0.02 &  0.15 &  2 & $-$0.23 &  0.20 &  7 & $-$0.27 &  0.21 &  7 \\
C1106-16402 & $-$0.21 &  0.16 & 26 & $-$0.23 &  0.16 & 20 &  0.24 &  0.07 &  3 & $-$0.04 &  0.14 &  8 & $-$0.16 &  0.16 &  8 &  0.07 &  0.14 &  2 & $-$0.28 &  0.19 &  7 & $-$0.33 &  0.20 &  7 \\
C1106-16451 &  0.12 &  0.20 & 25 &  0.11 &  0.19 & 19 &  0.65 &  0.03 &  3 &  0.05 &  0.11 &  8 & $-$0.07 &  0.12 &  8 &  0.01 &  0.15 &  2 & $-$0.27 &  0.19 &  5 & $-$0.32 &  0.20 &  5 \\
C1106-16566 & $-$0.22 &  0.13 & 25 & $-$0.23 &  0.13 & 20 &  0.18 &  0.10 &  2 & $-$0.04 &  0.10 &  8 & $-$0.16 &  0.13 &  8 & $-$0.00 &  0.18 &  2 & $-$0.27 &  0.19 &  8 & $-$0.32 &  0.20 &  8 \\
C1106-17160 & $-$0.62 &  0.12 & 25 & $-$0.64 &  0.15 & 19 & $-$0.35 &  0.08 &  3 & $-$0.32 &  0.12 &  8 & $-$0.46 &  0.15 &  8 & $-$0.28 &  0.09 &  2 & $-$0.41 &  0.12 & 10 & $-$0.48 &  0.14 & 10 \\
C1106-17565 & $-$0.11 &  0.20 & 25 & $-$0.13 &  0.20 & 19 &  0.65 &  0.02 &  3 &  0.04 &  0.12 &  9 & $-$0.10 &  0.13 &  9 &  0.09 &  0.10 &  2 & $-$0.35 &  0.19 &  7 & $-$0.40 &  0.20 &  7 \\
C1106-18086 & $-$0.69 &  0.07 & 24 & $-$0.70 &  0.09 & 18 & $-$0.47 &  0.11 &  3 & $-$0.30 &  0.08 &  7 & $-$0.46 &  0.14 &  7 & $-$0.41 &  0.12 &  2 & $-$0.60 &  0.10 &  9 & $-$0.69 &  0.14 &  9 \\
C1106-18101 & $-$0.44 &  0.09 & 27 & $-$0.45 &  0.09 & 21 & $-$0.23 &  0.04 &  3 & $-$0.28 &  0.10 &  7 & $-$0.40 &  0.17 &  7 & $-$0.14 &  0.14 &  2 & $-$0.42 &  0.24 & 11 & $-$0.47 &  0.25 & 11 \\
C1106-18303 & $-$0.05 &  0.15 & 25 & $-$0.06 &  0.15 & 19 &  0.52 &  0.05 &  3 &  0.13 &  0.11 &  8 &  0.01 &  0.10 &  8 &  0.07 &  0.10 &  2 & $-$0.29 &  0.20 &  7 & $-$0.34 &  0.21 &  7 \\
\hline
\end{tabular}
\tablefoot{$\sigma$ indicates the standard deviation in the line-to-line abundances. The Al abundance was derived only from the H-band spectral region.}
\end{table*}

\begin{table*}
\tiny
\setlength{\tabcolsep}{2.0pt}
\caption{Chemical abundance results for S, K, Ca, Ti, Cr, Ni, and Ce.}
\label{tab:abund2} 
\centering
\begin{tabular}{@{\extracolsep{2pt}}c r r r  r r r  r r r  r r r  r r r  r r r  r r r  r r r  }   
\hline\hline 
\multirow{2}{*}{ID}             & \multicolumn{3}{c}{S} & \multicolumn{3}{c}{K} & \multicolumn{3}{c}{Ca$_{\rm H-band}$}   & \multicolumn{3}{c}{Ti$_{\rm H-band}$} 
                                        & \multicolumn{3}{c}{Cr}        & \multicolumn{3}{c}{Cr$_{\rm NLTE}$}     & \multicolumn{3}{c}{Ni}        & \multicolumn{3}{c}{Ce} \\
\cline{2-4} \cline{5-7} \cline{8-10} \cline{11-13} \cline{14-16} \cline{17-19} \cline{20-22} \cline{23-25}
                                        & [X/H] & $\sigma$ & $N$        & [X/H] & $\sigma$ & $N$ & [X/H] & $\sigma$ & $N$ & [X/H] & $\sigma$ & $N$ 
                                        & [X/H] & $\sigma$ & $N$ & [X/H] & $\sigma$ & $N$ & [X/H] & $\sigma$ & $N$ & [X/H] & $\sigma$ & $N$ \\
\hline                           
C1103-7052 & $-$0.18 &  0.26 &  3 &  0.17 &  -- &  1 & $-$0.21 &  0.23 &  3 &  0.19 &  0.07 &  3 & $-$0.03 &  0.01 &  2 &  0.05 &  0.01 &  2 & $-$0.16 &  0.02 &  3 & $-$0.18 &  0.04 &  3 \\
C1103-8510 & $-$0.07 &  0.18 &  6 &  0.36 &  -- &  1 & $-$0.17 &  0.21 &  3 &  0.20 &  0.08 &  3 & $-$0.00 &  0.08 &  3 &  0.08 &  0.10 &  3 & $-$0.18 &  0.06 &  3 & $-$0.28 &  0.02 &  3 \\
C1103-9023 &  0.15 &  0.22 &  6 &  0.59 &  -- &  1 & $-$0.05 &  0.26 &  3 &  0.22 &  0.07 &  3 &  0.12 &  0.08 &  3 &  0.19 &  0.11 &  3 &  0.12 &  0.03 &  3 & $-$0.17 &  0.03 &  3 \\
C1103-9100 & $-$0.55 &  0.05 &  3 & $-$0.45 &  -- &  1 & $-$0.35 &  0.12 &  4 & $-$0.13 &  0.12 &  3 & $-$0.64 &  0.06 &  2 & $-$0.43 &  0.05 &  2 & $-$0.61 &  0.09 &  3 & $-$0.09 &  0.08 &  4 \\
C1103-9180 &  0.33 &  0.09 &  5 &  0.55 &  -- &  1 &  0.10 &  0.15 &  4 &  0.27 &  0.06 &  3 &  0.12 &  0.13 &  3 &  0.17 &  0.16 &  3 &  0.17 &  0.20 &  4 &  0.14 &  0.00 &  2 \\
C1103-9181 &  0.01 &  0.09 &  4 &  0.71 &  -- &  1 &  0.07 &  0.32 &  2 &  0.41 &  0.08 &  3 &  0.21 &  0.10 &  2 &  0.21 &  0.15 &  2 &  0.22 &  0.09 &  3 & $-$0.01 &  0.02 &  2 \\
C1103-9341 &  0.17 &  0.14 &  6 &  0.31 &  -- &  1 & $-$0.04 &  0.15 &  4 &  0.19 &  0.08 &  3 &  0.12 &  0.09 &  3 &  0.14 &  0.12 &  3 &  0.17 &  0.03 &  3 &  0.19 &  0.06 &  3 \\
C1103-9590 &  0.19 &  0.11 &  3 &  0.45 &  -- &  1 & $-$0.01 &  0.16 &  3 &  0.25 &  0.06 &  3 &  0.14 &  0.06 &  3 &  0.20 &  0.10 &  3 &  0.14 &  0.05 &  3 & $-$0.12 &  0.03 &  3 \\
C1103-10048 &  0.11 &  0.14 &  6 &  0.26 &  -- &  1 &  0.02 &  0.13 &  3 &  0.22 &  0.11 &  3 &  0.09 &  0.07 &  3 &  0.13 &  0.10 &  3 &  0.06 &  0.08 &  3 & $-$0.12 &  0.05 &  3 \\
C1103-10405 &  0.16 &  0.12 &  5 &  0.34 &  -- &  1 &  0.12 &  0.10 &  4 &  0.14 &  0.06 &  2 &  0.11 &  0.06 &  3 &  0.19 &  0.10 &  3 &  0.04 &  0.11 &  5 & $-$0.17 &  0.01 &  2 \\
C1103-10484 &  0.04 &  0.22 &  4 &  0.56 &  -- &  1 & $-$0.10 &  0.27 &  3 &  0.23 &  0.08 &  3 &  0.05 &  0.09 &  3 &  0.11 &  0.11 &  3 &  0.14 &  0.15 &  3 & $-$0.19 &  0.05 &  3 \\
C1103-11170 & $-$0.04 &  0.03 &  3 &  0.24 &  -- &  1 & $-$0.09 &  0.15 &  3 &  0.15 &  0.04 &  3 & $-$0.05 &  0.06 &  3 &  0.07 &  0.11 &  3 & $-$0.25 &  0.08 &  3 & $-$0.42 &  0.09 &  3 \\
C1103-11530 &  0.18 &  0.15 &  6 &  0.47 &  -- &  1 &  0.03 &  0.14 &  3 &  0.21 &  0.07 &  3 &  0.13 &  0.07 &  3 &  0.20 &  0.10 &  3 &  0.17 &  0.09 &  4 & $-$0.16 &  0.03 &  2 \\
C1103-12006 &  0.18 &  0.08 &  3 &  0.50 &  -- &  1 &  0.06 &  0.30 &  2 &  0.30 &  0.03 &  3 &  0.13 &  0.09 &  3 &  0.17 &  0.10 &  3 &  0.15 &  0.06 &  3 & $-$0.02 &  0.09 &  3 \\
C1103-12308 &  0.07 &  0.14 &  5 &  0.75 &  -- &  1 &  0.12 &  0.23 &  2 &  0.35 &  0.06 &  3 &  0.21 &  0.06 &  3 &  0.25 &  0.04 &  3 &  0.21 &  0.09 &  3 & $-$0.17 &  0.01 &  2 \\
\hline
C1106-15096 & $-$0.60 &  0.27 &  3 & $-$0.58 &  -- &  1 & $-$0.56 &  0.22 &  4 & $-$0.27 &  0.09 &  3 & $-$0.75 &  0.02 &  3 & $-$0.64 &  0.03 &  3 & $-$0.58 &  0.09 &  2 & $-$0.31 &  0.10 &  4 \\
C1106-15592 & $-$0.19 &  0.30 &  6 &  0.26 &  -- &  1 & $-$0.24 &  0.38 &  2 &  0.16 &  0.05 &  3 & $-$0.07 &  0.05 &  3 & $-$0.01 &  0.08 &  3 & $-$0.22 &  0.06 &  3 & $-$0.17 &  0.04 &  3 \\
C1106-16402 &  0.02 &  0.24 &  4 &  0.23 &  -- &  1 & $-$0.19 &  0.20 &  3 &  0.15 &  0.09 &  3 & $-$0.09 &  0.09 &  3 & $-$0.03 &  0.13 &  3 & $-$0.18 &  0.04 &  3 & $-$0.07 &  0.01 &  3 \\
C1106-16451 & $-$0.16 &  0.33 &  4 &  0.19 &  -- &  1 & $-$0.15 &  0.33 &  3 &  0.37 &  0.07 &  3 &  0.13 &  0.07 &  3 &  0.17 &  0.10 &  3 & $-$0.07 &  0.04 &  3 &  0.25 &  0.07 &  3 \\
C1106-16566 & $-$0.18 &  0.10 &  5 &  0.09 &  -- &  1 & $-$0.24 &  0.20 &  3 &  0.07 &  0.11 &  3 & $-$0.17 &  0.03 &  2 & $-$0.11 &  0.07 &  2 & $-$0.23 &  0.12 &  4 & $-$0.14 &  0.06 &  2 \\
C1106-17160 & $-$0.63 &  0.20 &  5 & $-$0.39 &  -- &  1 & $-$0.39 &  0.15 &  4 & $-$0.20 &  0.14 &  3 & $-$0.55 &  0.04 &  2 & $-$0.36 &  0.03 &  2 & $-$0.55 &  0.07 &  4 & $-$0.25 &  0.12 &  5 \\
C1106-17565 & $-$0.14 &  0.23 &  6 &  0.19 &  -- &  1 & $-$0.19 &  0.33 &  2 &  0.18 &  0.07 &  3 & $-$0.03 &  0.02 &  2 &  0.03 &  0.06 &  2 & $-$0.21 &  0.11 &  3 & $-$0.34 &  0.08 &  3 \\
C1106-18086 & $-$0.51 &  0.28 &  3 & $-$0.32 &  -- &  1 & $-$0.36 &  0.12 &  4 & $-$0.33 &  0.18 &  3 & $-$0.64 &  0.10 &  3 & $-$0.49 &  0.16 &  3 & $-$0.59 &  0.07 &  4 & $-$0.59 &  0.13 &  5 \\
C1106-18101 & $-$0.32 &  0.32 &  4 & $-$0.06 &  -- &  1 & $-$0.29 &  0.17 &  4 & $-$0.08 &  0.11 &  3 & $-$0.46 &  0.06 &  3 & $-$0.37 &  0.10 &  3 & $-$0.34 &  0.15 &  4 & $-$0.30 &  0.18 &  4 \\
C1106-18303 & $-$0.07 &  0.22 &  4 &  0.26 &  -- &  1 & $-$0.19 &  0.26 &  2 &  0.17 &  0.12 &  3 &  0.07 &  0.11 &  3 &  0.13 &  0.14 &  3 & $-$0.15 &  0.05 &  3 & $-$0.01 &  0.09 &  3 \\
\hline
\end{tabular}
\tablefoot{Ca and Ti abundances were derived only from the H-band spectral region.}
\end{table*}

\begin{table*}
\caption{Systematic error due to the uncertainty in the atmosphere parameters for C1103-10048.}
\label{tab:error} 
\setlength{\tabcolsep}{5pt}
\centering                                    
\begin{tabular}{@{\extracolsep{2pt}}c c c c c c c c c c} 
\hline\hline 
\multirow{2}{*}{Species}        & \multicolumn{2}{c}{$T_{\rm eff}$ (4119~K)}    & \multicolumn{2}{c}{$\log{g}$ (1.60~dex)}        & \multicolumn{2}{c}{[Fe/H] ($+$0.16~dex)}  & \multicolumn{2}{c}{$\xi_{t}$ (1.39~km~s$^{-1}$)}                              & \multirow{2}{*}{Total} \\
\cline{2-3} \cline{4-5} \cline{6-7} \cline{8-9} 
                                        & $-$86~K & $+$86~K                                     & $-$0.46~dex & $+$0.46~dex                       & $-$0.03~dex & $+$0.03~dex                             & $-$0.22~km~s$^{-1}$ & $+$0.22~km~s$^{-1}$       & \\
\hline
Fe      & $-$0.00       & $+$0.02       & $-$0.10       & $+$0.11       & $-$0.01 & $+$0.00 & $+$0.09 & $-$0.08 & 0.14 \\
Na      & $-$0.09 & $+$0.08 & $+$0.05 & $-$0.15 & $-$0.01 & $+$0.00 & $+$0.06 & $-$0.10 & 0.15 \\
Mg      & $-$0.04 & $+$0.02 & $-$0.04 & $-$0.03 & $-$0.02 & $+$0.00 & $+$0.05 & $-$0.06 & 0.07 \\
Al      & $-$0.08 & $+$0.08 & $-$0.08 & $+$0.02 & $+$0.00 & $-$0.00 & $+$0.10 & $-$0.07 & 0.13 \\
Si      & $-$0.03 & $-$0.05 & $-$0.21 & $+$0.12 & $+$0.00 & $-$0.01 & $+$0.06 & $-$0.11 & 0.19 \\
S       & $+$0.04 & $-$0.05 & $-$0.21 & $+$0.20 & $-$0.00 & $+$0.00 & $+$0.02 & $-$0.02 & 0.21 \\
K       & $-$0.04 & $+$0.05 & $-$0.09 & $+$0.12 & $+$0.02 & $-$0.01 & $+$0.08 & $-$0.06 & 0.13 \\
Ca      & $-$0.17 & $+$0.07 & $-$0.22 & $+$0.04 & $+$0.02 & $-$0.01 & $+$0.12 & $-$0.05 & 0.20 \\
Ti      & $-$0.09 & $+$0.09 & $-$0.05 & $+$0.04 & $+$0.00 & $+$0.00 & $+$0.06 & $-$0.04 & 0.11 \\
Cr      & $-$0.10 & $+$0.03 & $-$0.11 & $+$0.03 & $-$0.00 & $-$0.02 & $+$0.08 & $-$0.07 & 0.12 \\
Ni      & $+$0.00 & $-$0.00 & $-$0.11 & $+$0.10 & $-$0.01 & $+$0.01 & $+$0.06 & $-$0.05 & 0.12 \\
Ce      & $-$0.04 & $+$0.04 & $-$0.23 & $+$0.23 & $-$0.00 & $+$0.01 & $+$0.03 & $-$0.02 & 0.23 \\
\hline                                             
\end{tabular}
\end{table*}

We were also able to measure C, N, and O abundance from CO, CN and OH features, respectively. 
Because these elements are tied in molecules and therefore are  interrelated in spectral analysis,
we sequentially measured the O, C, and N abundances through spectrum synthesis and then iteratively performed these procedures while updating the CNO abundances.
However, CNO abundances are not included in this study due to the large line-to-line variation, and they are not essential for our purpose. 
We examined the effect on chemical abundance measurements for other elements by incorporating the CNO abundances in the model atmosphere. 
This effect is smaller than the typical measurement error because we measured abundances from the spectrum synthesis for the central region of each absorption line. 

In addition, for Fe, Mg, Si, and Cr, we estimated both LTE and non-LTE (NLTE) abundances using the line-by-line corrections  from the literature, where available, for spherical 1D MARCS models \citep{Bergemann2010, Bergemann2012, Bergemann2013, Bergemann2015}\footnote{http://nlte.mpia.de}. 
The lines used for the NLTE correction are marked in Table~\ref{tab:line}, and the NLTE abundances are listed in Tables~\ref{tab:abund} and \ref{tab:abund2}.
The NLTE corrections decrease the abundance of Fe ($\sim$0.02~dex), Mg ($\sim$0.11~dex) and Si ($\sim$0.06~dex) on average, whereas  the  Cr-abundance increases ($\sim$0.08~dex). 
We note that these studies also provide an NLTE correction for Ca and Ti, but our target stars are beyond the covered range for Ca, and half of the Ti lines are missing. 

The derived chemical abundances for each element ([X/H]), line-to-line scatter ($\sigma$), and the number of lines used (N) are listed in Tables~\ref{tab:abund} and \ref{tab:abund2}. We adopted the solar abundance scale of \citet{Asplund2009}.


\subsection{Error analysis} \label{sec:err}
We estimated the statistical errors for each element as $\sigma$/$\sqrt{N}$, which are marked in the figures. 
The mean error on the [Fe/H] measurements from all sample stars is 0.03~dex, which is comparable to those from high-resolution optical spectroscopy for stars in the bulge and GC \citep[$\sim$0.03~dex;][]{Munoz2018,Lim2021}, and from other NIR spectroscopy \citep[$\sim$0.10~dex;][]{Ishikawa2022}.
However, our uncertainties are larger than in the previous IGRINS study by \citet[][$<$ 0.01~dex]{Afsar2018} because their targets are far brighter than ours ($K$ $<$ 6.0) and have a high S/N ($>$ 200).

We also estimated uncertainties on the atmosphere parameters and the ensuing error on the abundances.
As discussed in Section~\ref{sec:param}, our determination of $T_{\rm eff}$ is based on the nine pairs of LDR estimate. 
We obtained uncertainties in $T_{\rm eff}$ from the standard error of the mean of these estimates for each star.
The median of these uncertainties for all target stars is 75~K, while the mean value is slightly higher at 80~K. 
We note, however, that the small scatter of the discrepancies between our $T_{\rm eff}$ estimates and those from EW$_{CO}$ and another LDR method by \citet{Jian2019} ($\lesssim$ 50~K; see Section~\ref{sec:teff} and Figure~\ref{fig:teff}) indicates that the actual uncertainties would be smaller than our estimates. 

The uncertainties in $\log{g}$ were computed by Monte Carlo sampling taking the uncertainty in $T_{\rm eff}$, random error of EW$_{CO}$, and fitting error of Eq~(\ref{eq:logg}) into account, while these are mainly driven by the uncertainty in $T_{\rm eff}$.
For example, a change of $\sim$200~K in $T_{\rm eff}$ causes a large variation in $\log{g}$ ($\sim$1.0~dex). 
This indicates that our $\log{g}$ determination is highly dependent on the estimate of $T_{\rm eff}$.
It is also worth noting that our derived $T_{\rm eff}$ and $\log{g}$ parameters are consistent with those of the APOGEE samples in the upper panel of Figure~\ref{fig:apo_param} despite these large uncertainties, although a direct star-by-star comparison could not be made.
In addition, we took the uncertainty in [Fe/H] from the standard error of the mean of the line-by-line Fe abundance measurements. 
The uncertainty in $\xi_{t}$ was then estimated by adopting all these uncertainties to the Eq~(\ref{eq:vt2}) through Monte Carlo sampling. 
Finally, the typical uncertainties of atmosphere parameter determinations for our samples are about 
$\Delta T_{\rm eff}$ = $\pm$75~K, 
$\Delta \log{g}$ = $\pm$0.43~dex,
$\Delta$[Fe/H] = $\pm$0.03~dex, 
and $\Delta \xi_{t}$ = $\pm$0.21~km~s$^{-1}$.
These uncertainties are somewhat larger than the general uncertainties of other spectroscopic studies because we employed empirical relations to derive atmosphere parameters.
We note that other IGRINS studies did not measure parameters from the NIR spectra, but rather used the literature  or optical data.

In order to examine the effect of these parameter uncertainties on the abundance measurement, we generated additional eight-atmosphere models with different parameters that we varied by their uncertainty for C1103-10048. 
We remeasured chemical abundances with these models from spectrum synthesis and then compared them with the original values. 
The differences in each abundance ratio depending on each parameter are listed in Table~\ref{tab:error}. 
Finally, an upper limit of the total systematic uncertainty by atmosphere parameter determinations was calculated as the squared sum of all contributions.

\subsection{Abundance differences between H-band and K-band spectral regions} \label{sec:HK}
\begin{figure}
\centering
   \includegraphics[width=0.32\textwidth]{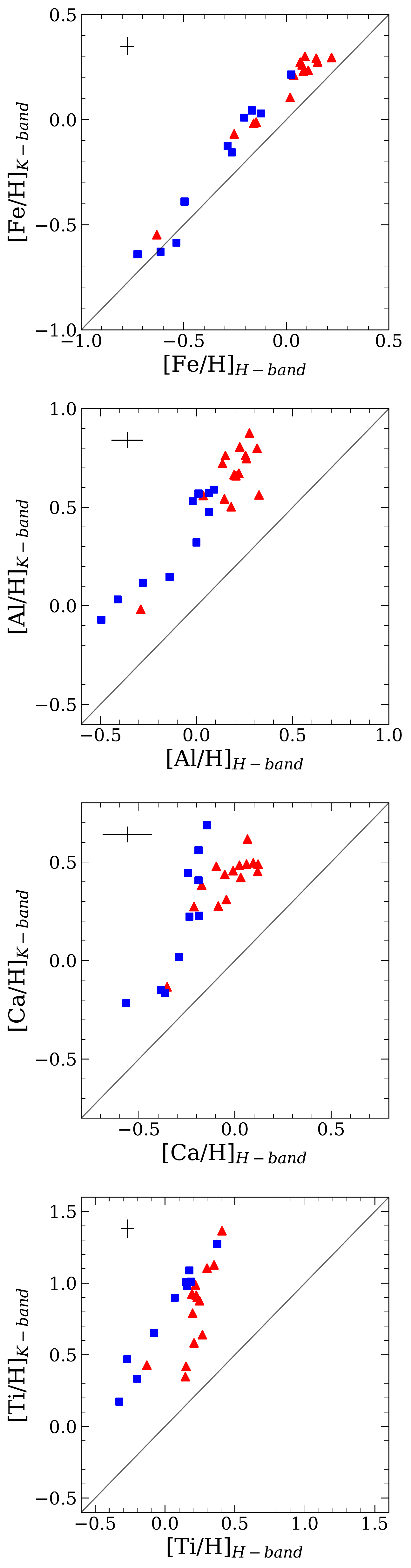}
     \caption{Comparison of [Fe/H], [Al/H], [Ca/H], and [Ti/H] abundance ratios derived from H-band and K-band spectral regions.  
      The symbols are the same as in Figure~\ref{fig:teff}, and the typical measurement error is plotted in the upper left corner of each panel. 
      The abundance ratios derived from the K band are generally higher than those from the H band. 
      These discrepancies are particularly significant for Al, Ca, and Ti elements. 
     }
     \label{fig:hk}
\end{figure}

We found systematic differences in the Al, Ca, and Ti abundances when comparing values from the H-band spectral regions with those from the K band.  
Figure~\ref{fig:hk} shows the comparison of [Fe/H], [Al/H], [Ca/H], and [Ti/H] when measured from the H- and K-band spectra.
The abundances measured from the K-band spectrum are generally higher than those from the H-band spectrum. 
In particular, the discrepancies of [Al/H], [Ca/H], and [Ti/H] are significant, with  mean differences of 0.46, 0.45, and 0.67~dex, which are much larger than their measurement errors (see Figure~\ref{fig:hk}). 
In the case of [Fe/H], although some difference is apparent ($\sim$ 0.14~dex), it is comparable to the typical 1$\sigma$ line-to-line scatter of the Fe measurements ($\sim$ 0.15~dex).
We also compared abundance ratios of Mg, Si, and S elements from the H- and K-band spectra. 
These elements show some discrepancy at the $<$ 0.2~dex level, but it is not as large as for the cases of Al, Ca, and Ti and is lower than the standard deviation of each element.  

One reason for these discrepancies could be erroneous atmosphere parameters because incorrectly measured $T_{\rm eff}$ and $\xi_{t}$ can cause a trend of abundance with wavelength. 
It is possible that our derived parameters are biased because our estimates are mainly based on the empirical relations from the literature.
However, these differences remain when we measured chemical abundances with altered atmosphere models in Section~\ref{sec:err}.
For instance, when we used a model at a higher gravity (1.60 $+$ 0.46~dex) for C1103-10048, the difference in abundances from H- and K-band spectra is slightly reduced from 0.46~dex to 0.39~dex for Al and 0.46~dex to 0.29~dex for Ca; however, the discrepancy for Ti remains, and the [Mg/Fe] ratio now differs.
Furthermore, when we adopt the stellar parameters from APOGEE for star C1106-18101 (the only one in common with our sample), the discrepancies between H- and K-spectra increase for Mg, Si, and Ca, while that in Ti is reduced.  
Therefore, if the atmosphere parameters are erroneous, this is not enough to cause this systematic discrepancy. 

On the other hand, the excitation potential and oscillator strength ($\log{gf}$) of the absorption line can affect the abundance measurements. 
We compared the line information used in this study with that from the recent National Institute of Standards and Technology (NIST) Atomic Spectra Database \citep{NIST_ASD} and the Kurucz line list.
The excitation potential is almost identical in these lists, while $\log{gf}$ is slightly different in the Kurucz list ($<$ 0.1~dex). 
When we adopted $\log{gf}$ values from the Kurucz list, the differences in abundance between H- and K-band spectra were still significant.
Therefore, it appears that the excitation potential and $\log{gf}$ of the lines are not the main reason.
One concern is that the excitation potentials of the used Ti lines are significantly different in the H- and K-band spectral regions, with lower values in the K band (see Table~\ref{tab:line}).
This discrepancy indicates that the sensitivity of the Ti abundance to T$_{eff}$ and its uncertainty could be different in the two spectral bands. 
In the case of Ca lines, the absorption lines in the K band have higher values of $\log{gf}$ than those in the H band, without significant differences in excitation potential. 
If the Ca lines in the K band form significantly deeper in the stellar atmosphere than other lines, such as the Fe lines we used to determine $\xi_t$, some discrepancies in the K-band Ca lines could be expected.
It is thus important to be aware of systematic differences in the characteristics of lines in different bands, especially when only a few lines are available.

\begin{figure}
\centering
   \includegraphics[width=0.4\textwidth]{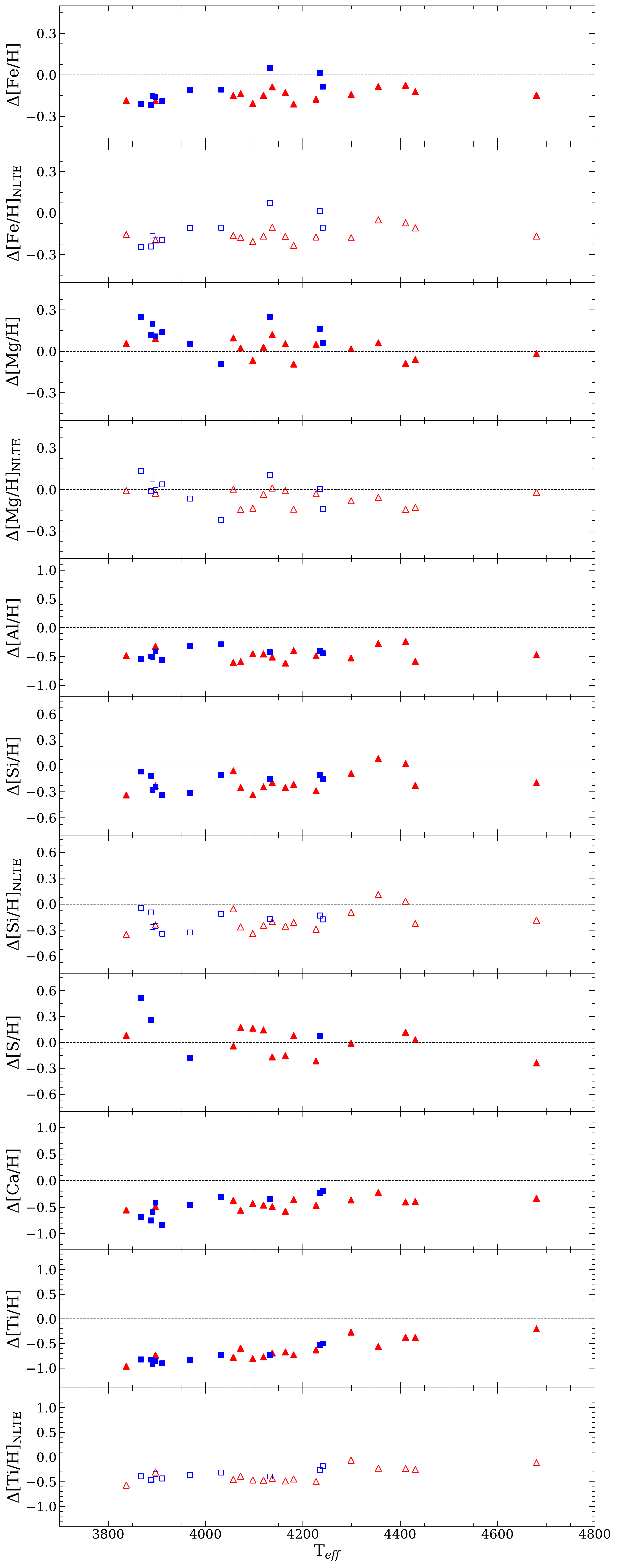}
     \caption{Difference in chemical abundance ratios of [Fe, Mg, Al, Si, S, Ca, and Ti/H] derived from H- and K-band spectra depending on $T_{\rm eff}$.
     The symbols are the same as in Figure~\ref{fig:teff}, and open symbols show the abundance ratios after NLTE correction.
     }
     \label{fig:HK_Teff}
\end{figure}

In Figure~\ref{fig:HK_Teff} we also examine the variation of abundance difference depending on $T_{\rm eff}$ for Fe, Mg, Al, Si, S, Ca, and Ti, where we were able to measure lines from the H- and K-band spectra. 
The differences in Fe, Si, Ca, and Ti abundances are correlated with $T_{\rm eff}$, while those of Mg and S are slightly anticorrelated. 
In most elements, the size of the differences decreases with increasing $T_{\rm eff}$.
This indicates that regardless of whether the derived $T_{\rm eff}$ is correct, the abundance discrepancy between the H  and K band is related with $T_{\rm eff}$ and is not a random variation. 
In this regard, the effect of NLTE on the abundance measurement might be one reason for the discrepancy between two spectral bands because the NLTE effects will differ for each element, absorption line, and star.  
As shown in Figure~\ref{fig:HK_Teff}, the trend of the abundance difference for [Mg/H] is reduced when we use the NLTE abundance, although this change is not noticeable for Fe and Si. 
The influence of the NLTE corrections is more evident in Ti, although we were able to apply the correction only for some of our lines. 
The average NLTE effect on the Ti abundance is 0.03~dex for H-band spectra, whereas it is $-$0.26~dex for K-band spectra. 
Thus, the typical abundance difference of [Ti/H] between H- and K-band spectra is reduced from 0.67 to 0.36 after NLTE correction. 
In particular, because these effects are larger at low $T_{\rm eff}$ stars, they can significantly reduce the trend with $T_{\rm eff}$ and also the abundance difference. 
For instance, the NLTE correction of Ti abundance in the K band is $-$0.52~dex for C1103-9181, which is the coolest star.
However, the abundance differences between H- and K-band spectra are still observed in the [Ti/H]$_{\rm NLTE}$ ratio (see the lower panel of Figure~~\ref{fig:HK_Teff}). 
We suspect that this is because the actual NLTE effect is larger than our estimate for these stars. 
In the same manner, the abundance differences in Al and Ca could be highly affected by NLTE effect for K-band spectra. 

In the upcoming section and Tables~\ref{tab:abund} and \ref{tab:abund2}, we only employed the H-band spectral region to measure abundances of Al, Ca, and Ti because the abundances measured from the K-band region are significantly overabundant with respect to the overall distribution of MW stars.
We note that these discrepancies have not been reported in previous IGRINS studies such as \citet{Afsar2018} and \citet{BocekTopcu2020}.

\section{Result} \label{sec:result}

\subsection{Metallicites and membership}
%
\begin{figure}
\centering
   \includegraphics[width=0.45\textwidth]{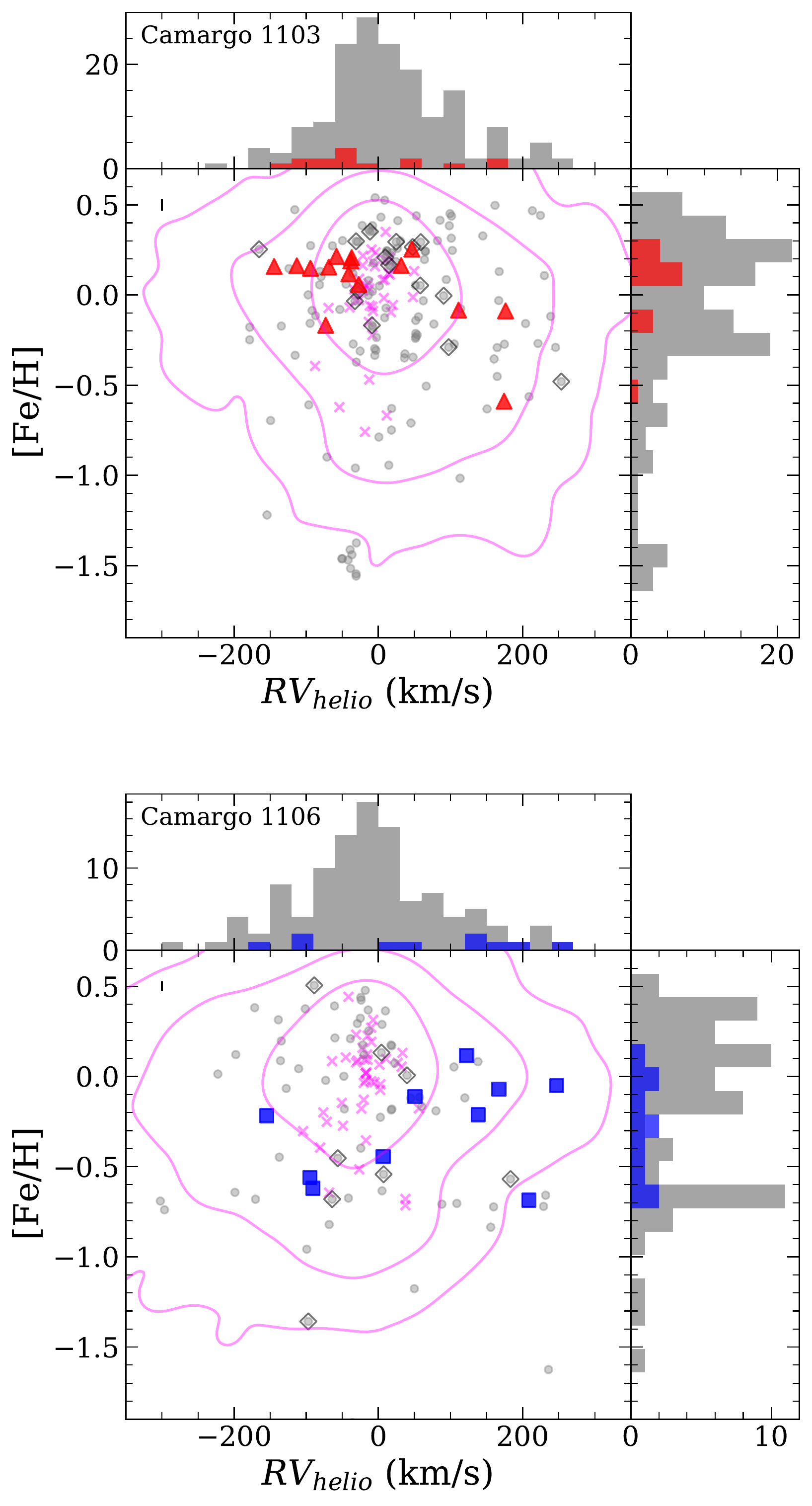}
     \caption{Distributions of our target stars on the [Fe/H] - RV$_{helio}$ plane, together with their histograms. 
     The symbols are the same as in Figure~\ref{fig:cmd}.
     The magenta lines show  1, 2, and 3$\sigma$ density contours  levels for the model stars from the Besan\c{c}on Galaxy model.
     Open diamonds and magenta X-symbols indicate giant stars located in the central region of cluster candidates ($<$ 5$\arcmin$) 
     within similar ranges of proper motions, magnitude, and colors as our targets from APOGEE and  Besan\c{c}on model, respectively. 
     The clump of APOGEE stars at [Fe/H] $\sim$ $-$1.5~dex in the upper panel is NGC~6544, which is a nearby foreground GC ($d$ $\sim$ 3.0~kpc). 
     }
     \label{fig:fe_rv}
\end{figure}
%
In order to examine the metallicity distribution of our target stars and to trace the evidence of metal-poor GCs, we show [Fe/H] versus RV$_{helio}$ in Figure~\ref{fig:fe_rv}.
The majority of stars located in the field of Camargo~1103 are more metal-rich than solar, while the metallicity range of stars located in the Camargo~1106 field varies from $-$0.7 to +0.2 dex. 
It is clear in both cases that there is no obvious overdensity in this parameter space that would unambiguously indicate the presence of a coherent, metal-poor stellar system as
advocated by \citet{Camargo2018}.
In the upper panel of Figure~\ref{fig:fe_rv}, it appears that six stars (C1103-9023, 9341, 9590, 10048, 10405, and 11530) form a clump at around [Fe/H] $\sim$ $+$0.16~dex and RV$_{helio}$ $\sim$ $-$45~km~s$^{-1}$. 
However, their metallicities are highly distinct from the expected value of a metal-poor GC \citep[$-$1.8 dex;][]{Camargo2018}, although it is possible that this is a metal-rich GC, but its metallicity is underdetermined from the isochrone fitting due to the uncertain reddening (see CMDs in Figure~\ref{fig:cmd}).
In addition, in the sense that the [Fe/H] and RV$_{helio}$ distributions of nearby APOGEE samples are similar to those of our targets, it is suspected that these stars are typical foreground disk stars or members of an open cluster. 
The distribution of our targets is more comparable to the APOGEE stars, which are located in the central region of Camargo~1103 ($<$ 5$\arcmin$) within similar selection criteria of magnitude, color, and proper motions as our targets (open diamonds in Figure~~\ref{fig:fe_rv}).
We also compared metallicity and radial velocity with the Besan\c{c}on Galaxy model\footnote{https://model.obs-besancon.fr/} \citep{Robin2003} for this field (within 10$\arcmin$ radius). 
The overall distribution of these model stars is similar to that of APOGEE stars, and their density peak is located near [M/H] $\sim$ $+$0.15~dex with RV$_{helio}$ $\sim$ $-$10~km~s$^{-1}$, covering the clump of our target stars (see contours in Figure~\ref{fig:fe_rv}). 
We note that giant stars of the Besan\c{c}on model located in the central region of Camargo~1103 with similar proper motions and magnitude to our targets (magenta X-symbols) are also mainly found in this peak locus.
In particular, the mean distance of the model stars in this locus on the metallicity-RV$_{helio}$ plane is estimated to be $\lesssim$3.5~kpc from the Sun. 
Thus, the observed clump containing six stars may be located in the foreground disk rather than the bulge.
The Gaia parallaxes of these stars, however, are highly uncertain to determine their distance (\texttt{parallax\_over\_error} $<$ 2.5).

In the case of Camargo~1106 field, we were unable to find any evidence of clustering among our target stars. 
The observed stars are widely spread in both [Fe/H] and RV$_{helio}$.
Unlike Camargo~1103, the distribution of stars in this field is somewhat different from that of the Besan\c{c}on Galaxy model. 
However, the distribution is quite similar to that of APOGEE samples, particularly with stars in the central region of Camargo~1106.
Although two stars (C1106-15096 and C1106-17160), which have similar [Fe/H] ($\sim-$0.6~dex) and RV$_{helio}$ ($\sim-$93~km~s$^{-1}$), might be GC stars, a sample size of two is still very small. 
Furthermore, this ``high'' metallicity is much higher than the estimate given by \citet{Camargo2018}. 
C1106-18086 ([Fe/H] = $-$0.69~dex and RV$_{helio}$ = 209~km~s$^{-1}$) is also peculiar because this star is far away from other observed stars and the Besan\c{c}on model in the parameter space shown in Figure~\ref{fig:fe_rv}.
However, because a number of APOGEE sample stars are also distributed in the range of $-$0.7 $<$ [Fe/H] $<$ $-$0.6 with various RV$_{helio}$, this star could originate from a common substructure with the APOGEE stars.

According to \citet{Camargo2018}, the stellar number density for Camargo~1103 and 1106 is estimated to be 15 $\sim$ 20 per arcmin$^{2}$ at R = 3$\arcmin$ and 2$\arcmin$ for the selected red giant stars. 
This number density indicates that more than 500 member stars are expected to be located in the inner 3$\arcmin$ radius region of Camargo~1103. 
Here, the 2MASS catalog includes 688 giant stars with $J$ $<$ 15 mag and 0.5 $<$ $J-H$ $<$ 2.0.
Thus, we can expect that roughly 70$\%$ of the stars in this inner region are GC members, although a number of stars could be undetected in the 2MASS catalog. 
On the other hand, as discussed in Section~\ref{sec:target}, we preselected 72 member candidates within selection criteria of magnitude, color, and proper motion located in a circle of 3$\arcmin$ radius from the center of Camargo 1103 (see the open red circles in Figure~\ref{fig:cmd}).  
If the 70$\%$  fraction of member stars is correct, these stars would be divided into 50 members and 22 nonmember stars.
Therefore, having observed 15 stars out of 72 candidates, it is hard to claim that all our targets were accidentally selected only from $\sim$22 nonmember stars. 
For Camargo~1106, a member fraction of 85$\%$ is derived from 210 expected members at r $<$ 2$\arcmin$ with an actual 245 stars in the 2MASS catalog, in the same manner as Camargo~1103. 
Thus, this fraction indicates that the 87 member candidates that satisfy our target selection criteria (open blue circles in Figure~\ref{fig:cmd}) consisted of 74 member and 13 nonmember stars. 
It is also unlikely that the observed 10 stars in the Camargo~1106 field are all nonmember stars.
Therefore, our results do not support the scenario that Camargo~1103 and 1106 are metal-poor GCs in the bulge. 

However, the reliability of the target selection can be a concern:
as described in Section~\ref{sec:target}, we selected targets as bright red giant stars from the 2MASS catalog, cross-matched with Gaia DR2. 
We note that \citet{Camargo2018} found Camargo~1103 and 1106 based on  decontaminated CMDs using 2MASS, WISE, and Gaia data. 
Because we employed not only the 2MASS CMD, but also Gaia photometry, mainly the relatively less reddened stars in the $J-K_{S}$ color were selected (see Figure~\ref{fig:cmd}). 
If Camargo~1103 and 1106 are more strongly affected by reddening, they would have much redder colors and fainter magnitudes than expected. 
In this case, it is possible that their true member stars were excluded by our CMD selection criteria.
In addition, it is also possible that cluster member stars are more centrally concentrated, while we selected targets within 3$\arcmin$ and 2$\arcmin$ from the center, but avoided the most crowded region.
We note, however, that \citet{Camargo2018} also selected cluster member stars within a radius of 3$\arcmin$ for Camargo~1103 and 2$\arcmin$ radius for Camargo~1106.
Thus, although we were unable to find evidence of any metal-poor GC from 15 and 10 stars in the Camargo~1103 and 1106 fields, respectively, there is still a low probability that we omitted cluster member stars by selection bias.

\subsection{Comparison with APOGEE}

\begin{figure*}
\centering
   \includegraphics[width=0.9\textwidth]{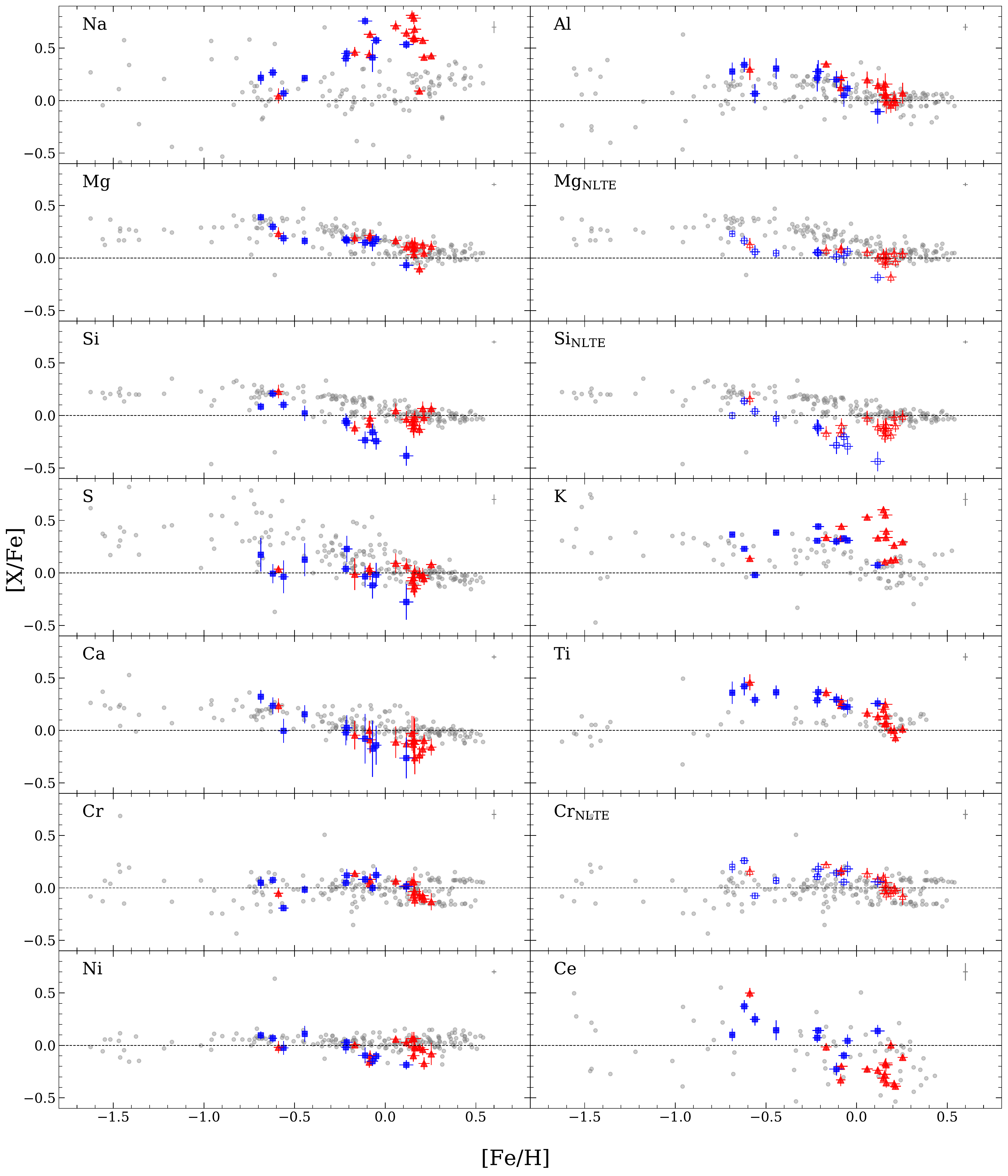}
     \caption{Chemical abundance ratios of our target stars, together with nearby APOGEE stars ($d$ $<$ 15$\arcmin$). 
     The symbols are the same as in Figure~\ref{fig:cmd}, and open symbols show abundance ratios after NLTE correction. 
     The vertical and horizontal bars for each marker indicate the measurement error for each element, which is taken as $\sigma$/$\sqrt{N}$ from Tables~\ref{tab:abund} and \ref{tab:abund2}. 
     The gray cross in the upper right corner of each panel indicates the typical measurement error of the APOGEE samples. 
     }
     \label{fig:abund}
\end{figure*}

In Figure~\ref{fig:abund} we show the chemical abundances of 11 elements (Na, Mg, Al, Si, S, K, Ca, Ti, Cr, Ni, and Ce) in comparison with the nearby APOGEE stars. 
The overall abundance distributions for our targets are fairly comparable to those of APOGEE samples, which is consistent with the scenario that the  atmosphere parameters of our targets also chiefly overlap with APOGEE (see Figure~\ref{fig:apo_param}). 
In more detail, a number of stars with [Fe/H] $>$ $-$0.3~dex are enhanced in [Na/Fe]  compared to the APOGEE stars. 
This is a useful feature because an Na-enhancement of stars is usually observed in present-day GCs or in stars with an origin in GCs \citep[see, e.g.,][]{Carretta2009, Lim2021}. 
However, further data are required because our Na abundance is only measured in the K-band spectral region.  
As we discussed in Section~\ref{sec:HK}, the chemical abundances of Al, Ca, and Ti obtained from the K-band spectrum are generally more enhanced than those from the H-band spectrum. 
If this situation also affected Na abundance measurement, it is possible that our [Na/Fe] ratios are systematically higher than those of APOGEE.
However, we were unable to derive Na abundances from the lines in H-band spectra, which are used in APOGEE \citep{Smith2021}, because of the severe blending and the lower sensitivity to the Na abundance of these lines.

The [Mg/Fe] ratio is one of the key tracers of an accretion origin, with Mg being an $\alpha$-element. The abundance ratios of our stars agree well with the trend of APOGEE with a small measurement error. 
Several stars are relatively depleted in [Mg/Fe] abundances compared to other stars with similar [Fe/H].
This result demonstrates that we can trace the accretion origin of stars using the IGRINS data, although two of them are metal-rich stars with [Fe/H] $>$ 0.0~dex (C1103-9341 and C1106-16451), which means that there is little possibility of an accretion origin. 
With the NLTE abundance ratio ([Mg/Fe]$_{\rm NLTE}$), however, our measurements follow the lower boundary of the APOGEE sample, where we note that the abundances of the APOGEE are shifted in zeropoint without any NLTE correction \citep[see][]{Jonsson2020}.

The abundances of the other $\alpha$-elements, Si, S, and Ca, are slightly lower than  the APOGEE stars at any given metallicity, although the intrinsic trends with [Fe/H] are consistent. 
Similar situations are also shown in K and Ti  in that our data are somewhat enhanced in [K/Fe] for stars with [Fe/H] $>$ 0.0~dex and in [Ti/Fe] for stars with [Fe/H] $<$ 0.0~dex than APOGEE sample.
Because their trends with [Fe/H] are approximately consistent, these discrepancies would be systematic and not random. 
It is important to note that we applied NLTE corrections only for Fe, Mg, Si, and Cr, but these effects are non-negligible and are different for each element and each star. 
In particular, because the NLTE effects are correlated with metallicity, this correction could either strengthen or reduce the abundance trends with metallicity for some elements. 
Therefore, while we can examine the chemical properties of stars from our own data, careful and equivalent consideration of the NLTE corrections for both our observation and the reference data is required for a meaningful comparison \citep[see, e.g.,][]{Zhang2016, Osorio2020}.


\section{Discussion} \label{sec:discussion}
With the aim of confirming and characterizing two low-latitude star cluster candidates, we obtained NIR spectra using the IGRINS spectrographs at the Gemini-South telescope of stars toward the Galactic bulge. 
The derived stellar parameters agree well with those of nearby APOGEE stars, while some chemical abundances show some systematic differences in several elements. 
Most strikingly, we were unable to find any evidence of metal-poor GCs in either of the purported Camargo~1103 and 1106 fields from the metallicity versus RV plane. 
Thus, it appears that there are no GCs around these fields in our data, although we cannot completely rule out a low probability that we have omitted member stars from the observation.

As described in Section~\ref{sec:HK}, we found that some abundance ratios measured from H- and K-band spectra show significant and systematic discrepancies, particularly for Al, Ca, and Ti.  
Although we were unable to clarify the origin of these systematic differences, the effects of atmosphere parameters, 
atomic parameters (excitation potential and $\log{gf}$), or NLTE are suspected.
We have also shown that these abundance differences are reduced, even though not eliminated, after NLTE correction (see Figure~\ref{fig:HK_Teff}).
\citet{Masseron2021} reported that the 1D NLTE correction is qualitatively identical but quantitatively different from the 3D NLTE model. In this regard, the remaining abundance differences after NLTE correction may be due to the limitation of the current 1D NLTE model. 
Consequently, if the NLTE effect causes these large and systematic differences, it indicates that more careful and detailed NLTE corrections are essential for NIR spectroscopy. 
Furthermore, the extended 3D NLTE model will become more important in stellar chemical abundance studies \citep[see also][]{Masseron2021}.
On the other hand, it is also possible that the incorrectly defined continuum for the K-band spectrum for some reasons, such as inaccurate telluric correction, contamination by molecular bands, or effect of line blanketing, causes this discrepancy in the abundances, although we were unable to examine this. 
We note that because of the difficulty of continuum level determination for metal-rich stars, we adjusted the continuum level for each absorption line after continuum normalization for entire H- and K-band spectra, respectively. 

Although no evidence of metal-poor GCs is detected among our targets, we were able to derive atmosphere parameters and chemical abundances for stars toward the Galactic bulge from IGRINS NIR spectroscopy. 
Our study indicates that spectroscopic follow-up is required to confirm the newly reported GC candidates in the bulge. 
We expect that the recent precise Gaia data and statistical test for the distribution of stars will be helpful for a more efficient high-resolution study of high-probability cluster members.
With a more careful target selection procedure and some further calibrations for chemical abundance, it is clear that the observation using IGRINS is useful to examine the detailed chemical properties of stars and clusters in the bulge. 
We will continue our observation for the bulge region to better understand the formation and evolution of the MW.


\vspace{5mm}
\begin{acknowledgements}
We thank the referee for a number of helpful suggestions.
DL and AJKH gratefully acknowledge funding by the Deutsche Forschungsgemeinschaft (DFG, German Research Foundation) -- Project-ID 138713538 -- SFB 881 (``The Milky Way System''), subprojects A03, A05, A11. 
SHC acknowledges support from the National Research Foundation of Korea (NRF) grant funded by the Korea government (MSIT) (NRF-2021R1C1C2003511).
YWL and SH acknowledge support from the National Research Foundation of Korea (2022R1A2C3002992, 2022R1A6A1A03053472).
DL thanks Sree Oh for the consistent support.
This work used the Immersion Grating Infrared Spectrometer (IGRINS) that was developed under a collaboration between the University of Texas at Austin and the Korea Astronomy and Space Science Institute (KASI) with the financial support of the Mt. Cuba Astronomical Foundation, of the US National Science Foundation under grants AST-1229522 and AST-1702267, of the McDonald Observatory of the University of Texas at Austin, of the Korean GMT Project of KASI, and Gemini Observatory.
This research made use of Astropy,\footnote{http://www.astropy.org} a community-developed core Python package for Astronomy \citep{AstropyCollaboration2013, AstropyCollaboration2018}.

\end{acknowledgements}


\bibliographystyle{aa} 
\bibliography{export-bibtex} 

\end{document}